\documentclass[11pt]{article}
\usepackage[a0paper]{geometry}
\usepackage{anysize}
\marginsize{3cm}{3cm}{2cm}{2cm}
\usepackage{amsmath}
\usepackage{amsfonts}
\usepackage{bbm}
\usepackage{amssymb}
\usepackage{slashed}
\usepackage{mathrsfs}
\usepackage{graphicx}
\usepackage{epsfig}
\usepackage{enumerate}
\usepackage{dsfont}
\usepackage{appendix}
\usepackage{pdfpages}
\usepackage{setspace}
\usepackage{wrapfig}
\usepackage[utf8]{inputenc}
\usepackage{bm}
\usepackage{braket}
\usepackage{float}
\usepackage{epstopdf} 
\usepackage{authblk}
\usepackage{multicol}
\usepackage{enumitem}
\usepackage{bbold}
\usepackage{subcaption}
\usepackage[normalem]{ulem}

\newcommand{\e}{\text{e}}

\newcommand{\bph}{\bm{\phi}}

\title{Post-selected flavor entanglement in pion–pion scattering}

\date{}

\numberwithin{equation}{section}

\author[1]{Victor Miguel Banda Guzm\'an \thanks{victor.banda@upslp.edu.mx} }

\author[2]{Rub\'en Flores-Mendieta
\thanks{ruben@ifisica.uaslp.mx} }

\author[2]{Johann Hern\'andez
\thanks{johann.edir.hy@gmail.com}}

\affil[1]{Universidad Polit\'ecnica de San Luis Potos\'i, Urbano Villal\'on 500, Col. La Ladrillera, C.P. 78363, San Luis Potos\'i, S.L.P., M\'exico.}

\affil[2]{Instituto de F\'isica, Universidad Aut\'onoma de San Luis Potos\'i, Alvaro Obreg\'on 64, Zona Centro, San Luis Potos\'i, S.L.P. 78000, M\'exico.}

\begin{document}
\maketitle

\begin{abstract}

Quantum entanglement provides a quantitative measure of nonclassical correlations and plays a central role in quantum information theory. Identifying and quantifying the mechanisms by which fundamental interactions generate or suppress entanglement is therefore of conceptual interest. In this work, flavor entanglement generated in pion-pion scattering is analyzed within a post-selected framework, in which the outgoing two-pion state is conditioned on fixed asymptotic momenta. The analysis is carried out in the isospin-symmetric limit using chiral perturbation theory, including one-loop corrections to the scattering amplitude. A general formalism is developed to quantify post-selected flavor entanglement through the von Neumann entropy of the reduced flavor density matrix, expressed in terms of a flavor tensor characterizing the initial state. This framework is applied to initially unentangled charged two-pion states. Except for the $\ket{++}$ and $\ket{--}$ channels, which remain unentangled due to isospin conservation, the post-selected final states exhibit nontrivial flavor entanglement and behave effectively as entangled qubits or qutrits. The entanglement is maximal near threshold and typically peaks at scattering angles close to $\theta=\pi/2$. The analysis demonstrates that isospin-channel dominance plays a central role in shaping the entanglement structure: the $I=0$ channel drives generic initial states toward highly entangled qutrit configurations after scattering and post-selection. Conversely, suitably chosen initial superpositions can undergo a reduction of entanglement, showing that the strong interaction can act both as a generator and a suppressor of quantum correlations. One-loop corrections quantitatively redistribute entanglement across phase space, sharpening angular structures that appear more diffuse at tree level.

\end{abstract}

\section{Introduction}\label{Introduction}

Two fundamental features distinguishing quantum systems from their classical counterparts are superposition and entanglement. When two quantum systems are entangled, information encoded in their degrees of freedom becomes intrinsically correlated, even if they are spatially separated by large distances. Understanding the mechanisms by which entanglement is generated or suppressed by physical interactions is therefore of central importance, not only for quantum information but also for the foundations of quantum field theory.

In the context of the strong interaction, recent work has revealed a nontrivial interplay between quantum entanglement and emergent symmetries in low-energy hadronic processes. Within a unitary $S$-matrix framework, it has been shown that the suppression of entanglement correlates with the appearance of approximate spin-flavor symmetries in baryon-baryon scattering, suggesting that reduced entanglement may be a characteristic feature of the infrared regime of QCD \cite{beane, Low:2021, Low:2023}. Flavor entanglement has been investigated in this context for a variety of systems \cite{ Kowalska:2024, Thaler}, including pion scattering \cite{Beane:2021}. 

An alternative perspective is offered by post-selection frameworks in which the final state is conditioned on specific measurement outcomes, such as fixed outgoing momenta after scattering \cite{pesch,faleiro,Low:2024,Low:2024_2,Cervera-Lierta,Fedida,Blasone:2024,Blasone:2024_2}. Unlike the full $S$-matrix approach, which sums over all asymptotic channels including forward scattering events that are operationally equivalent to free propagation, the post-selection framework isolates genuinely interacting processes. This distinction allows one to probe how entanglement is generated or suppressed specifically by the interaction itself, rather than being averaged out by noninteracting contributions.

Motivated by this conceptual difference, the present work employs a post-selection approach to investigate flavor entanglement in pion-pion scattering at one-loop order in chiral perturbation theory. This framework complements existing $S$-matrix-based analyses by addressing a distinct and experimentally meaningful notion of entanglement: the entanglement present in detected scattering events conditioned on momentum transfer. In doing so, it provides insight into how low-energy QCD dynamics can act both as a mechanism for either the generation or suppression of quantum entanglement, depending on the initial pion-flavor state. In this sense, the two approaches probe distinct notions of entanglement in scattering processes.

This article is organized as follows. Section \ref{sec:CHPT} summarizes the salient aspects of chiral perturbation theory needed for the analysis in order to set notation and conventions. Section~\ref{sec:pionpionscat} presents the one-loop pion-pion scattering amplitude in the isospin-symmetric limit, expressed in terms of the physical pion mass $M_\pi$, the decay constant $F_\pi$, and the low-energy constants appearing at order $\mathcal{O}(\frac{1}{F_\pi^4})$. This section includes density plots of the relative magnitude of the $I=0,1,2$ isospin channels and provides a qualitative discussion of flavor entanglement generation for initially unentangled charged pions.

Section \ref{sec:Postselected_formalism} presents a rigorous formulation of the post-selection framework based on a wave-packet analysis and introduces the von Neumann entropy as the primary tool to quantify flavor entanglement generated or suppressed in pion-pion scattering.

Section~\ref{sec:entangpions} contains the main results of this work. A quantitative analysis of the entanglement produced by scattering and post-selection with fixed momenta is carried out for initially unentangled charged pion states. A similar analysis is performed for the post-selection suppression of the entanglement of initially entangled states. This section concludes with a discussion of tree-level versus one-loop entropy calculations. The conclusions are summarized in Sec.~\ref{sec:Conclusions}.

For completeness and reproducibility, the manuscript is supplemented by several appendices that detail intermediate steps on pion-pion scattering, often omitted in the literature. Appendix~\ref{App:ScatteringAmplitudeOneloop} provides a detailed derivation of the one-loop pion-pion scattering amplitude from the quantum effective action, reproducing the classic result of Gasser and Leutwyler~\cite{gasser}. Appendix~\ref{app:FTidentityL24} presents a useful identity for evaluating the one-loop pion-pion scattering amplitude. Appendix~\ref{app:feynmanint} collects definitions and identities for the one-loop integrals appearing in the calculation. Appendix~\ref{app:oneloopmr} contains the one-loop computation of the pion mass and the residue of the exact pion propagator. Appendix~\ref{app:WavePack} summarizes the identities used in the wave-packet formulation of the post-selection framework, and Appendix~\ref{app:SphericalBasis} presents the spherical isospin basis $\ket{I,M}$ in terms of Cartesian pion states and derives the corresponding von Neumann entropies.

\section{\label{sec:CHPT}Chiral perturbation theory at order $\mathcal{O}(p^4)$}

Chiral perturbation theory (ChPT) provides an effective field-theory description of low-energy QCD based on the spontaneous breaking of the chiral symmetry
$\mathrm{SU}(2)_L \times \mathrm{SU}(2)_R \to \mathrm{SU}(2)_V$.
The associated Goldstone bosons are identified with the pions and are described by the matrix field
\begin{equation}
U(x)=\exp\!\left(\frac{i\,\phi(x)}{F}\right), \label{eq:defMspace}
\end{equation}
where $F$ denotes the pion bare decay constant and $\phi(x)=\sum_{a}\tau^a\phi^a(x)$, with $\tau^a$ representing the Pauli matrices and $\phi^a(x)$ real scalar fields for flavor $a$. The effective Lagrangian is constructed using a nonlinear realization of the chiral group, ensuring invariance under $\mathrm{SU}(2)_L \times \mathrm{SU}(2)_R$
\cite{scherer2,scherer1,weinberg,salam,coleman,callan2,brauner}.

Up to next-to-leading order in the chiral expansion, the effective Lagrangian reads
\begin{equation}
\mathcal{L}=\mathcal{L}_2+\mathcal{L}_4, \label{eq:ChPTLag}
\end{equation}
where the leading-order contribution is given by
\begin{equation}
\mathcal{L}_2=\frac{F^2}{4}\, \label{eq:Lag2}
\mathrm{Tr}\!\left(D_\mu U D^\mu U^\dagger\right)
+\frac{F^2}{4}\,
\mathrm{Tr}\!\left(\chi U^\dagger+U\chi^\dagger\right),
\end{equation}
and the next-to-leading-order term reads
\begin{equation}
\mathcal{L}_4=\sum_{i=1}^7 l_i\,L_i
+\sum_{i=1}^3 h_i\,H_i, \label{eq:Lag4}
\end{equation}
Here, $L_i$ and $H_i$ denote the standard $\mathcal{O}(p^4)$ operators of chiral perturbation theory, while $l_i$ and $h_i$ are its corresponding low-energy constants \cite{scherer2,scherer1,gasser,maiani}.

The explicit form of the operators, the definition of the covariant derivative, and the external sources are summarized in Appendix~\ref{App:ScatteringAmplitudeOneloop}.

\section{\label{sec:pionpionscat}Pion-pion scattering}

This section summarizes the structure of pion-pion scattering within chiral perturbation theory, with the specific aim of identifying the dynamical ingredients relevant for the generation of flavor entanglement. The results presented here are standard; however, their organization in terms of isospin channels will play a central role in the post-selected entanglement analysis developed in the following sections.

Let $T$ denote the interacting part of the $S$-matrix,
\begin{equation}
S = 1 + iT .
\label{S_matrix_splitting}
\end{equation}
For the scattering process $\phi^a(p_1)+\phi^b(p_2)\to\phi^c(p_3)+\phi^d(p_4)$, the $T$-matrix elements can be parametrized as \cite{scherer1},
\begin{eqnarray}
\braket{p_4,d;p_3,c|T|p_1,a;p_2,b}
&=& (2\pi)^4 \delta\!\left(\sum_{i=1}^4 p_i\right)
T^{abcd}(p_1,p_2,p_3,p_4)
\nonumber\\
&& \hspace{-120pt} = (2\pi)^4 \delta\!\left(\sum_{i=1}^4 p_i\right)
\Big[
\delta^{ab}\delta^{cd} A(s,t,u)
+\delta^{ac}\delta^{bd} A(t,s,u)
+\delta^{ad}\delta^{bc} A(u,t,s)
\Big],
\label{eq:Tmtxelem}
\end{eqnarray}
where the function $A(s,t,u)$ satisfies the crossing relation $A(s,t,u)=A(s,u,t)$.

The Mandelstam variables $s$, $t$ and $u$ are defined as\footnote{All external momenta are taken to be incoming.}
\begin{subequations}
\label{eq:mandelstam}
\begin{eqnarray}
s &=& (p_1+p_2)^2=(p_3+p_4)^2,\\
t &=& (p_1+p_3)^2=(p_2+p_4)^2,\\
u &=& (p_1+p_4)^2=(p_2+p_3)^2, , \label{Mandelstam}
\end{eqnarray}
\end{subequations}
and satisfy the constraint
\begin{equation}
s+t+u=4M_\pi^2,
\label{Maldestam_identity}
\end{equation}
with $M_\pi$ the pion mass in the isospin-symmetric limit.

The scattering amplitude $A(s,t,u)$ can be computed systematically in chiral perturbation theory. At next-to-leading order, $\mathcal{O}(p^4)$, it receives contributions from tree-level terms in $\mathcal{L}_4$ and from one-loop diagrams generated by $\mathcal{L}_2$. For completeness, a detailed derivation is provided in Appendix~\ref{App:ScatteringAmplitudeOneloop}, where the one-loop Gasser-Leutwyler expression for $A(s,t,u)$ from Ref. \cite{gasser} is reproduced in Eq. \eqref{eq:Aoneloop}.

The one-loop expression for $A(s,t,u)$ is naturally written in terms of the bare parameters $M$ and $F$ of the chiral Lagrangian. For numerical applications and physical interpretation, it is convenient to express the amplitude in terms of the physical pion mass $M_\pi$ and decay constant $F_\pi$.

Using the one-loop relations between bare and physical parameters \cite{gasser}, one finds
\begin{subequations}
\label{Aprox_MF}
\begin{equation}
M = M_\pi \left(1+\frac{M_\pi^2}{64\pi^2F_\pi^2}\bar{l}_3\right),
\end{equation}
\begin{equation}
F = F_\pi \left(1-\frac{M_\pi^2}{16\pi^2F_\pi^2}\bar{l}_4\right),
\end{equation}
\end{subequations}
where $\bar l_3$ and $\bar l_4$ are renormalized scale-independent low-energy constants.

Substituting Eq.~\eqref{Aprox_MF} into the one-loop amplitude yields, up to $\mathcal{O}(F_\pi^{-4})$,
\begin{eqnarray}
A(s,t,u) & = & \frac{s-M_\pi^2}{F_\pi^2} + \frac{1}{6F_\pi^4} \Bigg\{ \left[14M_\pi^4+t(s+2t)-2M_{\pi}^2(2s+5t) \right] \bar{J}_2(t) \nonumber \\
&& + \left[14M_\pi^4+u(s+2u)-2M_\pi^2(2s+5u) \right] \bar{J}_2(u) + 3(s^2-M_\pi^4) \bar{J}_2(s) \Bigg\} \nonumber \\
& & \mbox{}  + \frac{1}{96 \pi^2 F_\pi^4} \Bigg\{ 2 \bar{l}_1 \left(s-2 M_\pi^2\right)^2 +2 \bar{l}_2 \left(-8 M_\pi^4+4 M_\pi^2 s+t^2+u^2\right) \nonumber \\
& & \mbox{}  + \frac13 \left[53 M_\pi^4-24 M_\pi^2 s-8 s^2-5 \left(t^2+u^2\right)\right] \Bigg\} + \dfrac{M_\pi^2}{32 \pi^2 F_\pi^4} \bigg[ 4 s \bar{l}_4 \nonumber \\
&& - M_\pi^2 \left(\bar{l}_3+4 \bar{l}_4\right) \bigg]. \label{eq:AoneloopPhysical}
\end{eqnarray}
This expression has the same functional structure as the original one-loop result, with the bare parameters replaced by their physical counterparts and additional finite terms accounting for this redefinition at the given order.

It is often convenient to rewrite the amplitude in a basis of definite total isospin. The $T$-matrix can be decomposed as
\begin{equation}
T^{abcd}=\sum_{I=0}^2 A_I(s,t,u)\,
\left[\mathcal{P}_F^{(I)}\right]^{abcd},
\label{eq:AlternativeTmatrix}
\end{equation}
where $A_I(s,t,u)$ are the amplitudes in the $I=0,1,2$ channels and $\mathcal{P}_F^{(I)}$ are the corresponding flavor projectors. The explicit expressions for the amplitudes and projectors are given by,
\begin{eqnarray}
    A_0(s,t,u) &=& 3 A(s,t,u) + A(t,s,u) + A(u,t,s), \nonumber \\
    A_1(s,t,u) &=& A(t,s,u) - A(u,t,s), \nonumber \\
    A_2(s,t,u) &=& A(t,s,u) + A(u,t,s). \label{def:ATotalIsospin}
\end{eqnarray}
and
\begin{eqnarray}
    \left[\mathcal{P}^{(0)}_F\right]^{abcd} &=& \dfrac{1}{3} \delta^{ab} \delta^{cd}, \nonumber \\
    \left[\mathcal{P}^{(1)}_F\right]^{abcd} &=& \dfrac{1}{2} \left( \delta^{ac} \delta^{bd} - \delta^{ad} \delta^{bc} \right), \nonumber \\
    \left[\mathcal{P}^{(2)}_F\right]^{abcd} &=& \dfrac{1}{2} \left( \delta^{ac} \delta^{bd} + \delta^{ad} \delta^{bc} \right) - \dfrac{1}{3} \delta^{ab} \delta^{cd}. \label{def:FlavorProjectors}
\end{eqnarray}

To illustrate the relative importance of the different isospin channels, Fig.~\ref{Plot:relative_mag_amplitudes} displays density plots of the relative magnitude
\begin{equation}
r_I(s,t,u)=\frac{|A_I(s,t,u)|}{\sum_{J=0}^2|A_J(s,t,u)|},\qquad I=0,1,2.
\label{def:relative_magnitude}
\end{equation}
The numerical evaluation is performed in the center-of-mass frame, with
\begin{equation}
s=4E^2,\qquad
t=-\frac{s-4M_\pi^2}{2}(1-\cos\theta),\qquad
u=-\frac{s-4M_\pi^2}{2}(1+\cos\theta),
\end{equation}
using the physical values $M_\pi=139~\mathrm{MeV}$, $F_\pi=92~\mathrm{MeV}$ and the low-energy constants $\bar l_1=-0.4$, $\bar l_2=4.3$, $\bar l_3=3.0$, $\bar l_4=4.4$ \cite{LECs}.

Figure~\ref{Plot:relative_mag_amplitudes} shows that the isoscalar channel $I=0$ provides the dominant contribution within the kinematical region displayed in the density plots, accounting for approximately $65\%$--$85\%$ of the total amplitude. The $I=2$ channel gives the next-largest contribution, while the $I=1$ channel is generally subdominant, becoming relevant only at higher energies and at forward or backward scattering angles.

This hierarchy among isospin channels will be shown to have direct consequences for the generation of flavor entanglement in pion-pion scattering. This is illustrated qualitatively in the following section taking as an example initially unentangled charged pion states.  

\subsection{Qualitative description of flavor entanglement for initially charged pion states}
\label{qualitative_des_charged}

The charged pion states $\ket{\pi^+}$, $\ket{\pi^-}$, and $\ket{\pi^0}$ can be expressed in terms of the Cartesian isospin basis $\ket{\phi^a}$ as
\begin{subequations}
\begin{eqnarray}
\ket{\pi^0} & = & \ket{\phi^3}, \\
\ket{\pi^+} & = & \frac{1}{\sqrt{2}} \left( \ket{\phi^1}-i\ket{\phi^2} \right), \\
\ket{\pi^-} & = & \frac{1}{\sqrt{2}} \left( \ket{\phi^1}+i\ket{\phi^2} \right). \label{eq:physicalpions}
\end{eqnarray}
\end{subequations}
For notational convenience, the physical pion states $\ket{\pi^+}$, $\ket{\pi^-}$, and $\ket{\pi^0}$ will be denoted simply by $\ket{+}$, $\ket{-}$, and $\ket{0}$, respectively. 

Initial two-pion flavor states are conveniently expanded in the basis of total isospin $\ket{I,M}$ as
\begin{align}
\ket{++} &= \ket{2,+2}, &\quad
\ket{--} &= \ket{2,-2}, \nonumber\\
\ket{+0} &=
\frac{1}{\sqrt{2}}\ket{2,+1}
+\frac{1}{\sqrt{2}}\ket{1,+1}, &\quad 
\ket{0+} &=
\frac{1}{\sqrt{2}}\ket{2,+1}
-\frac{1}{\sqrt{2}}\ket{1,+1}, \nonumber \\
\ket{-0} &=
\frac{1}{\sqrt{2}}\ket{2,-1}
+\frac{1}{\sqrt{2}}\ket{1,-1}, &\quad
\ket{0-} &=
\frac{1}{\sqrt{2}}\ket{2,-1}
-\frac{1}{\sqrt{2}}\ket{1,-1}, \nonumber \\
\ket{+-} &=
\frac{1}{\sqrt{6}}\ket{2,0}
+\frac{1}{\sqrt{2}}\ket{1,0}
+\frac{1}{\sqrt{3}}\ket{0,0}, &\quad 
\ket{-+} &=
\frac{1}{\sqrt{6}}\ket{2,0}
-\frac{1}{\sqrt{2}}\ket{1,0}
+\frac{1}{\sqrt{3}}\ket{0,0}, \nonumber \\
\ket{00} &=
\sqrt{\frac{2}{3}}\ket{2,0}
-\frac{1}{\sqrt{3}}\ket{0,0}. \label{def:chargedtwopions}
\end{align}

\begin{figure}[h]
\scalebox{0.35}{\includegraphics{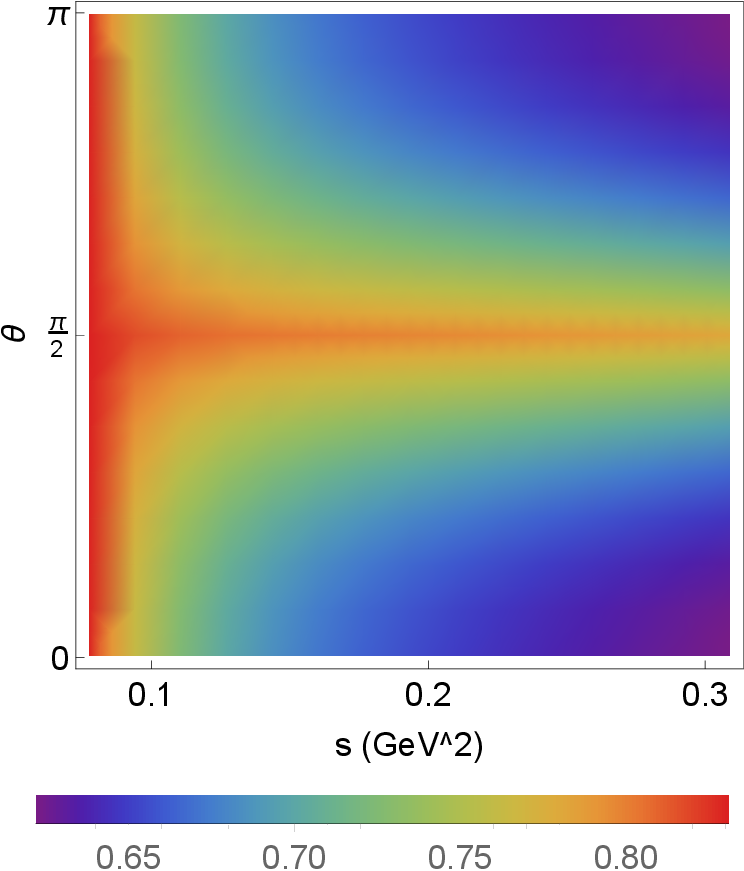}}
\scalebox{0.35}{\includegraphics{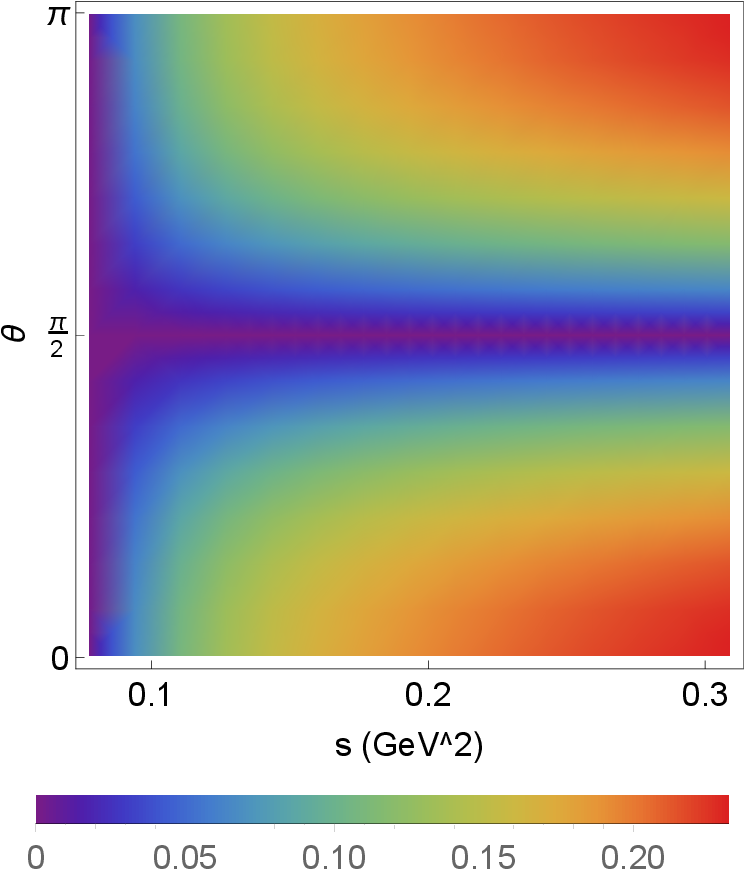}}
\scalebox{0.35}{\includegraphics{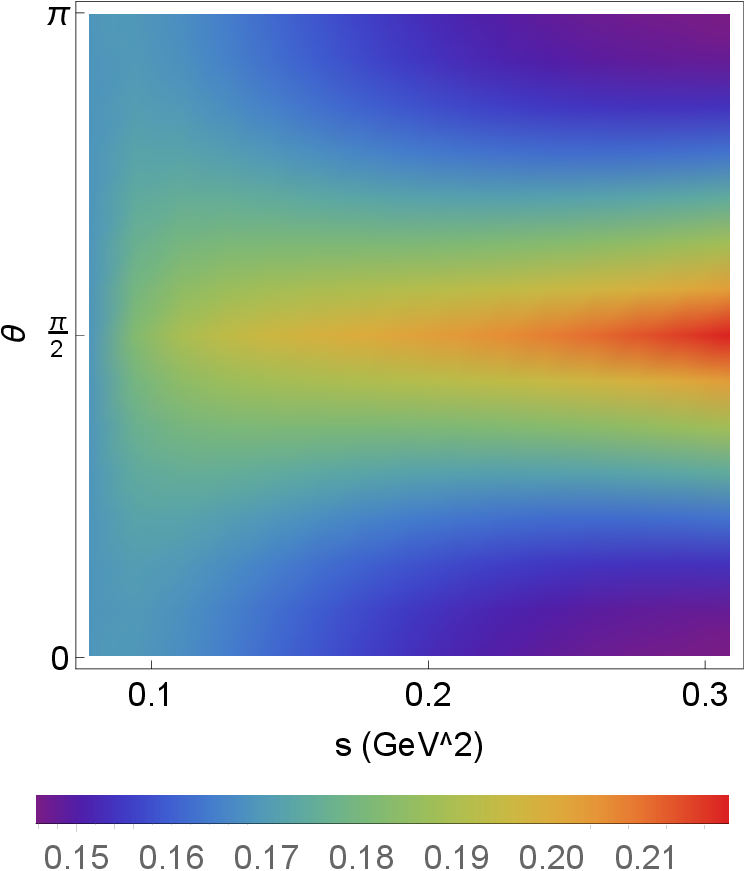}}
\caption{\label{Plot:relative_mag_amplitudes}Density plots for the relative magnitude amplitude $r_0(s,t,u)$ (left), $r_1(s,t,u)$ (center) and $r_2(s,t,u)$ (right) defined in Eq.~\eqref{def:relative_magnitude} as functions of the Mandelstam variable $s$ and the scattering angle $\theta$ in the center-of-mass frame.}
\end{figure}

From Fig.~\ref{Plot:relative_mag_amplitudes}, Table \ref{table:SphericalBasisVNE}, and the isospin decomposition above, the following qualitative picture emerges:
\begin{enumerate}
\item The states $\ket{++}$ and $\ket{--}$ belong to the $I=2$ sector and therefore remain unchanged by the scattering process. As a result, they remain unentangled both before and after the interaction.

\item Since the $I=0$ channel dominates the scattering amplitude, states with a sizable isoscalar component, such as $\ket{+-}$, $\ket{-+}$ and $\ket{00}$, are expected to evolve into highly entangled qutrit flavor configurations.

\item For states such as $\ket{+0}$, $\ket{0+}$, $\ket{-0}$ and $\ket{0-}$, the larger weight of the $I=2$ channel relative to $I=1$ suggests entanglement properties closer to those of the $\ket{2,\pm1}$ sector.
\end{enumerate}

While the discussion above provides useful intuition, a quantitative definition of entanglement in scattering processes requires conditioning on detected momenta. This motivates the post-selected density-matrix framework developed in the next section.

\section{\label{sec:Postselected_formalism}Post-selected formalism for flavor entanglement}

This section develops the theoretical framework used to quantify flavor entanglement generated by pion-pion scattering within a post-selected description of the scattering process. The central goal is to isolate entanglement that is genuinely produced by strong interactions, rather than entanglement that arises trivially from the structure of asymptotic scattering states in a full $S$-matrix treatment.

In conventional $S$-matrix–based analyses, entanglement measures are constructed from asymptotic in- and out-states that include forward-propagating components corresponding to noninteracting evolution. While this approach is well defined and unitary, it effectively averages over interacting and noninteracting events, thereby obscuring the entanglement specifically generated by momentum exchange in a scattering process. From an operational perspective, such forward events are indistinguishable from free propagation and would not contribute to entanglement observed in post-selected scattering experiments.

To address this issue, we adopt a post-selection framework in which the final state is conditioned on detecting outgoing particles with specified momenta different from those of the incoming wave packets. This procedure excludes forward scattering contributions and isolates genuinely interacting events. The resulting notion of entanglement is therefore conditional and operational: it quantifies the quantum correlations present in detected scattering outcomes rather than in the full asymptotic state. Similar post-selected approaches have been successfully employed in recent studies of entanglement in relativistic scattering processes~\cite{pesch,faleiro,Low:2024,Low:2024_2,Cervera-Lierta,Fedida,Blasone:2024,Blasone:2024_2}.

This post-selected notion of entanglement is complementary to, rather than in conflict with, entanglement measures defined at the level of the full $S$-matrix.

\subsection{Entanglement measure}
\label{subsec:ent_measure}

The amount of entanglement generated in the scattering process is quantified using the quantum mutual information. For a bipartite system composed of subsystems $A$ and $B$, the mutual information is defined as \cite{desur}
\begin{equation}
S(\rho_A,\rho_B) = S(\rho_A) + S(\rho_B) - S(\rho_{AB}),
\label{mutual_information}
\end{equation}
where $\rho_{AB}$ denotes the density matrix of the full system and $S(\rho)$ is the von Neumann entropy,
\begin{equation}
S(\rho) = - \mathrm{Tr}(\rho \log \rho).
\label{eq:VNentropygral}
\end{equation}
with $\mathrm{Tr}$ denoting the trace operation and logarithms are taken to base~2.

If $\lambda_i$ are the eigenvalues of $\rho$, the entropy can equivalently be written as
\begin{equation}
S(\rho)=-\sum_i\lambda_i\log\lambda_i,
\label{eq:VNentropyeigen}
\end{equation}
with the convention $0\log0=0$.

If $\rho_{AB}$ in Eq.~\eqref{mutual_information} is obtained from a final pure state produced by a $2 \rightarrow 2$ scattering process, the mutual information for the two particles $A$ and $B$ in the scattering event simplifies to,
\begin{equation}
S(\rho_A,\rho_B)=2S(\rho_A)=2S(\rho_B),
\label{eq:mutual}
\end{equation}
so that the von Neumann entropy of either reduced density matrix provides a complete characterization of the bipartite entanglement. Although the von Neumann entropy is not an entanglement monotone for generic mixed states, it constitutes a faithful entanglement measure for pure bipartite systems, which is the relevant situation after conditioning on a definite scattering outcome. This choice therefore suffices for the operational notion of entanglement considered here.

\subsection{Initial-state wave-packet formulation}
\label{subsec:wavepacket}

The initial two-pion state is taken to be
\begin{eqnarray}
\ket{i;a,b}=\sum_{a',b'}\int d^3\bar{k}_1\,d^3\bar{k}_2\,
\phi_{\bm{p}_1}(\bm{k}_1)\phi_{\bm{p}_2}(\bm{k}_2)
U^{ab a'b'}\ket{k_1,a'}\ket{k_2,b'},
\label{eq:initalstate}
\end{eqnarray}
where $U^{ab a'b'}$ is a tensor operator acting in flavor space. The Lorentz-invariant integration measure is defined as
\begin{equation}
d^3\bar{k}=\frac{d^3k}{(2\pi)^3 2E_{\bm{k}}}.
\end{equation}

The functions $\phi_{\bm{p}}(\bm{k})$ are on-shell Gaussian wave packets centered around momentum $\bm{p}$ with width $\sigma$\footnote{All the wave packets are assumed to have the same width for simplicity.}~\cite{Giunti},
\begin{equation}
\phi_{\bm{p}}(\bm{k})=
\left(\frac{2\pi}{\sigma^2}\right)^{3/4}
(2E_{\bm{k}})^{1/2}
\exp\!\left[-\frac{(\bm{k}-\bm{p})^2}{4\sigma^2}\right].
\label{eq:wavepacket}
\end{equation}

The introduction of wave packets serves two purposes. First, it regularizes the momentum-conserving delta functions appearing in $S$-matrix elements, rendering all expressions finite. Second, it provides an operational description of scattering events with finite experimental resolution. In the limit $\sigma\to0$, the wave packets become sharply localized in momentum space.

With this normalization, a single-particle wave-packet state
\begin{equation}
\ket{\psi_{\bm{p}}}=\int d^3\bar{k}\,\phi_{\bm{p}}(\bm{k})\ket{k}
\end{equation}
satisfies $\langle\psi_{\bm{p}}|\psi_{\bm{p}}\rangle=1$, assuming the relativistic normalization
\begin{equation}
\langle p|q\rangle=(2\pi)^3 2E_{\bm{p}}\delta^3(\bm{p}-\bm{q}).
\end{equation}
Moreover, Gaussian integration yields
\begin{equation}
\langle\psi_{\bm{p}}|\psi_{\bm{p}'}\rangle=
\exp\!\left[-\frac{(\bm{p}-\bm{p}')^2}{8\sigma^2}\right],
\label{eq:orthowavepacket}
\end{equation}
which vanishes for $\bm{p}\neq\bm{p}'$ in the sharp-packet limit, ensuring effective orthogonality of distinct momentum states.

\subsection{Post-selected final state and reduced flavor density matrix}
\label{subsec:postselection}

The outgoing state after scattering is obtained by acting with the $S$-matrix on the initial state,
\begin{eqnarray}
\ket{f;a,b}&=&S\ket{i;a,b}\nonumber\\
&=&\ket{i;a,b}
+i\sum_{a',b'}\int d^3\bar{k}_1\,d^3\bar{k}_2\,
\phi_{\bm{p}_1}(\bm{k}_1)\phi_{\bm{p}_2}(\bm{k}_2)
U^{ab a'b'}T\ket{k_1,a'}\ket{k_2,b'}.
\label{eq:finalstate}
\end{eqnarray}

To isolate entanglement generated by genuine interactions, a post-selection condition is imposed on the outgoing momenta. Specifically, non-forward scattering events are selected using the projection operator
\begin{equation}
\Pi_{p_3,p_4}=
\ket{\psi_{\bm{p}_3},\psi_{\bm{p}_4}}
\bra{\psi_{\bm{p}_3},\psi_{\bm{p}_4}}
\otimes\mathbf{1}_F,
\label{def:PorjectionoperatorMomentum}
\end{equation}
with $\bm{p}_3\neq\bm{p}_1$ and $\bm{p}_4\neq\bm{p}_2$. This procedure excludes events that are experimentally indistinguishable from free propagation and ensures that the resulting entanglement originates from the scattering interaction itself.

The density matrix after unitary evolution is
\begin{equation}
\rho_f=\ket{f;a,b}\bra{f;b,a},
\end{equation}
and the post-measurement density matrix is defined as
\begin{equation}
\rho_{\mathrm{PM}}=
\frac{\Pi_{p_3,p_4}\rho_f\Pi_{p_3,p_4}^\dagger}
{\mathrm{Tr}\!\left(\Pi_{p_3,p_4}\rho_f\Pi_{p_3,p_4}^\dagger\right)}.
\label{eq:rhoPM}
\end{equation}

Tracing over momentum degrees of freedom yields the flavor density matrix
\begin{equation}
\rho_F=\int d^3\bar{p}\,d^3\bar{q}\,
\langle p,q|\rho_{\mathrm{PM}}|p,q\rangle.
\label{def:reducedflavorrho}
\end{equation}

Since $\braket{\psi_{\bm{p}_4},\psi_{\bm{p}_3}|\psi_{\bm{p}_3},\psi_{\bm{p}_4}} = 1$, the expression for $\rho_F$ in Eq. \eqref{def:reducedflavorrho} becomes,
\begin{eqnarray}
    \rho_F = \dfrac{1}{\text{Tr}\left(\Pi_{p_3,p_4} \rho_f \Pi^\dagger_{p_3,p_4}\right)} \sum_{c,d,c^\prime,d^\prime} \braket{\psi_{\bm{p}_4}\psi_{\bm{p}_3};c^\prime, d^\prime|f;a,b} \braket{f;b,a|\psi_{\bm{p}_3},\psi_{\bm{p}_4};c,d} \ket{c^\prime,d^\prime} \bra{c,d}, \nonumber \\ \label{eq:rhof_2}
\end{eqnarray}
where 
\begin{eqnarray}
    \braket{\psi_{\bm{p}_4}\psi_{\bm{p}_3};c^\prime, d^\prime|f;a,b} = \braket{\psi_{\bm{p}_4}\psi_{\bm{p}_3};c^\prime, d^\prime|i} + i \sum_{a^\prime b^\prime} U^{aba^\prime b^\prime} M^{a^\prime b^\prime c^\prime d^\prime}, \label{def:productWavePacketf}
\end{eqnarray}
with $M^{a b c d}$ defined as
\begin{eqnarray}
    M^{a b c d} = \int \left( \prod_{i=1}^4 d^3 \bar{k}_i \right) \left( \prod_{i=1}^4 \phi_{\bm{p}_i}(\bm{k}_i) \right) \braket{k_3,k_4;c,d|T|k_1,k_2;a,b}. \label{def:Mamplitude}
\end{eqnarray}

By the orthogonality of the wave packets in the limit $\sigma \to 0$, the first term in Eq.~\eqref{def:productWavePacketf} vanishes. The second term can then be evaluated to leading order in $\sigma$ using Eq.~\eqref{eq:Tmtxelem} together with the wave–packet result given in Eq.~\eqref{eq:Isigma_result}. Upon substituting these into Eq.~\eqref{def:productWavePacketf}, the equation can be rewritten as,
\begin{eqnarray}
    \braket{\psi_{\bm{p}_4}\psi_{\bm{p}_3};c^\prime, d^\prime|f;a,b} = \beta\, R^{abc^\prime d^\prime}, \label{eq:prod_wavepacketfinal}
\end{eqnarray}
where
\begin{eqnarray}
    \beta = \dfrac{(2\pi)^{11}\sigma^2}{\left(\prod_{i=1}^4 E_{\bm{p}_i}\right)\sqrt{\alpha \det \mathcal{Q} }},
\end{eqnarray}
with $\alpha$ and $Q$ defined in equations \eqref{def:alphaV} and \eqref{def:mathcalQ}, respectively, and
\begin{eqnarray}
    R^{abc^\prime d^\prime} = A(s,t,u)\, \delta^{c^\prime d^\prime} \sum_{h} U^{abhh} + A(t,s,u) U^{abc^\prime d^\prime} + A(u,t,s) U^{ab d^\prime c^\prime}. \label{def:RTensor}
\end{eqnarray}

Using the result in Eq.~\eqref{eq:prod_wavepacketfinal}, and after performing the trace operation in the numerator of Eq.~\eqref{eq:rhof_2}, the latter equation can be rewritten as,
\begin{eqnarray}
    \rho_F = \dfrac{1}{N_{ab}} \sum_{c,d,c^\prime,d^\prime} R^{abc^\prime d^\prime} R^{\ast \,abc d} \ket{c^\prime,d^\prime} \bra{c,d}, \label{eq:rhoF}
\end{eqnarray}
where the normalization factor $N_{ab}$ is given by,
\begin{eqnarray}
    N_{ab} &=& \Big[2 \,\text{Re}\left\{ A(s,t,u) A^\ast(t,s,u) \right\} + 2 \,\text{Re}\left\{ A(s,t,u) A^\ast(u,t,s) \right\} + \nonumber \\
    &&  + 3|A(s,t,u)|^2 \Big] \left| \sum_{h} U^{abhh} \right|^2 + \left( |A(t,s,u)|^2+|A(u,t,s)|^2 \right) \sum_{c,d} |U^{abcd}|^2 \nonumber \\
    && + 2 \text{Re}\left\{ A(t,s,u) A^\ast(u,t,s) \sum_{c,d} U^{abcd} \left(U^{abdc}\right)^\ast \right\}. \label{def:Nab}
\end{eqnarray}

Finally, tracing over one of the pion yields
\begin{equation}
\rho_{\mathrm{red}}=
\frac{1}{N_{ab}}
\sum_{c,d,c'}
R^{ab c'd}R^{*ab cd}
\ket{c'}\bra{c},
\label{eq:rhoF_reduced}
\end{equation}
whose von Neumann entropy quantifies the flavor entanglement generated by the scattering process.

\section{\label{sec:entangpions}Entanglement generation and suppression in pion--pion scattering from post-selection}

This section applies the post-selected framework introduced in Sec.~\ref{sec:Postselected_formalism} to examine how strong-interaction dynamics can both generate and suppress flavor entanglement in pion--pion scattering. By considering specific classes of initial two-pion states, the mechanisms responsible for entanglement enhancement are identified, as well as the conditions under which scattering and post-selection lead to partial or complete disentanglement. The discussion emphasizes the physical interpretation of the results, highlighting the role of isospin decomposition, kinematical constraints, and loop-level corrections in shaping the entanglement structure of the post-selected final states.

\subsection{Quantitative flavor entanglement for initially unentangled charged pion states}
\label{Sec:entropy_chargedstates}

The quantitative generation of flavor entanglement is first analyzed for the initially flavor-unentangled charged-pion states defined in Eq.~\eqref{def:chargedtwopions}. These states are reproduced within the wave-packet formalism of Sec.~\ref{sec:Postselected_formalism} by choosing the flavor tensor operator $U^{abcd}$ in Eq.~\eqref{eq:initalstate} as
\begin{eqnarray}
    U^{abcd} = C^{ac} C^{bd},
    \label{def:UCharged}
\end{eqnarray}
where $C^{ab}$ is a $3\times3$ matrix defined in the Cartesian basis,
\begin{eqnarray}
    C^{ab} =
    \begin{pmatrix}
        0 & 0 & 1 \\
        \dfrac{1}{\sqrt{2}} & -\dfrac{i}{\sqrt{2}} & 0 \\
        \dfrac{1}{\sqrt{2}} & \dfrac{i}{\sqrt{2}} & 0
    \end{pmatrix}.
    \label{def:C_tensor}
\end{eqnarray}
Acting on the Cartesian field vector $(\phi^1,\phi^2,\phi^3)$, the matrix $C$ generates the physical pion fields $(\pi^0,\pi^+,\pi^-)$.

Substituting Eq.~\eqref{def:UCharged} into Eqs.~\eqref{def:Nab} and \eqref{eq:rhoF_reduced}, the von Neumann entropies obtained after scattering and post-selection for the initially unentangled charged-pion states take the explicit form
\begin{align}
    S_{++} & = 0, \nonumber \\
    S_{+0} &=
    \frac{|A_t|^2}{|A_t|^2+|A_u|^2}
    \log\!\left(\frac{|A_t|^2+|A_u|^2}{|A_t|^2}\right)
    + \frac{|A_u|^2}{|A_t|^2+|A_u|^2}
    \log\!\left(\frac{|A_t|^2+|A_u|^2}{|A_u|^2}\right), \nonumber \\
    S_{00} &=
    2\frac{|A_s|^2}{N_{00}}
    \log\!\left(\frac{N_{00}}{|A_s|^2}\right)
    + \frac{|\bar{A}|^2}{N_{00}}
    \log\!\left(\frac{N_{00}}{|\bar{A}|^2}\right), \nonumber \\
    S_{+-} &=
    \frac{|A_s|^2}{N_{+-}}
    \log\!\left(\frac{N_{+-}}{|A_s|^2}\right)
    + \frac{|A_s+A_t|^2}{N_{+-}}
    \log\!\left(\frac{N_{+-}}{|A_s+A_t|^2}\right)
    + \frac{|A_s+A_u|^2}{N_{+-}}
    \log\!\left(\frac{N_{+-}}{|A_s+A_u|^2}\right),
    \label{eq_VNE_charged}
\end{align}
where the subscripts on $S$ label the initial charged–pion states and
\begin{eqnarray}
    N_{00} &=& |\bar{A}|^2 + 2|A_s|^2, \nonumber \\
    N_{+-} &=& \dfrac{1}{3} |A_0|^2 + \dfrac{1}{2} |A_1|^2 + \dfrac{1}{6} |A_2|^2, \nonumber \\
    \bar{A}(s,t,u) &=& A_s + A_t + A_u.
    \label{def_BarNA}
\end{eqnarray}
The shorthand notation
\begin{eqnarray}
    A_s &=& A(s,t,u), \nonumber \\
    A_t &=& A(t,s,u), \nonumber \\
    A_u &=& A(u,t,s),
    \label{def:shorthandA}
\end{eqnarray}
has been adopted for compactness.

The remaining entropies follow directly from isospin symmetry and satisfy
\begin{align}
    S_{++} = S_{--}, \qquad
    S_{+0} = S_{-0} = S_{0+} = S_{0-}, \qquad
    S_{+-} = S_{-+}.
    \label{eq:VNE_charged_rest}
\end{align}

As a direct consequence of isospin invariance of the $S$--matrix, the charged flavor states $\ket{++}$ and $\ket{--}$ remain unentangled under scattering and post-selection, which is reflected in their identically vanishing von Neumann entropies.

At threshold, defined by $s = 4M_\pi^2$ and $t=u=0$, the von Neumann entropies in Eq.~\eqref{eq_VNE_charged}, evaluated at next-to-leading order in the chiral expansion, reduce to
\begin{eqnarray}
    S_{+0} &=& 1, \nonumber \\
    S_{00} &=& \frac{1}{361} \left[ 361 \log (19) -684 \log (3) + 
    \left(16 \bar{l}_1+32 \bar{l}_2-6 \bar{l}_3+21\right)\,\frac{ M_\pi^2\, \log(3) }{\pi^2 F_\pi^2} \right], \nonumber \\
    S_{+-} &=& \dfrac{1}{289} \left[289 \log (17)-34 \log (5038848) + (16 \bar{l}_1+32 \bar{l}_2-6 \bar{l}_3+21) \frac{M_\pi^2 \log \left(\frac{3}{2}\right)}{\pi ^2 F_{\pi}^2} \right] . \nonumber \\
    \label{eq:VN_charged_threshold}
\end{eqnarray}

Equation~\eqref{eq:VN_charged_threshold} shows that at threshold the entropy $S_{+0}$ is independent of the low-energy constants and saturates the maximal von Neumann entropy of an entangled qubit, $S_{+0}=1$. This result follows from the equal weighting of the $t$- and $u$-channel amplitudes at threshold.

Using the physical inputs $M_\pi = 139\,\mathrm{MeV}$ and $F_\pi = 92\,\mathrm{MeV}$ together with the low-energy constants $\bar{l}_1 = -0.4$, $\bar{l}_2 = 4.3$, and $\bar{l}_3 = 3.0$, the remaining threshold entropies evaluate numerically to
\begin{eqnarray}
    S_{00} = 1.38, \qquad S_{+-} = 1.53. \label{eq:VN_charged_threshold_numerical}
\end{eqnarray}
These values lie close to the maximal entropy $\log(3)$ associated with a maximally entangled qutrit, indicating that post-selected pion scattering can generate near-maximal flavor entanglement in multi-level systems.

\begin{figure}[h]
\scalebox{0.35}{\includegraphics{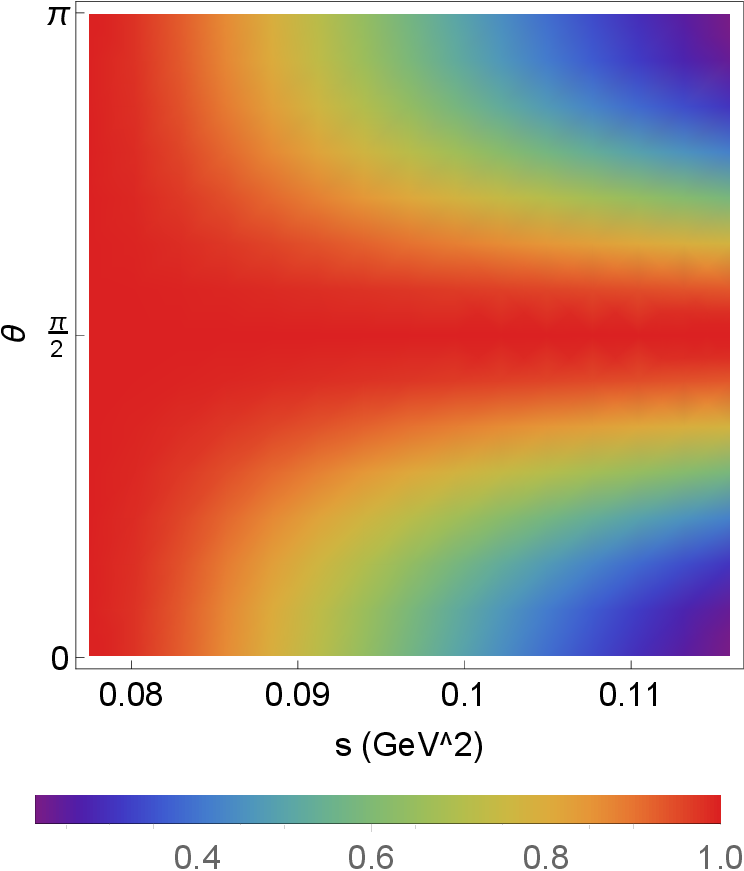}}
\scalebox{0.35}{\includegraphics{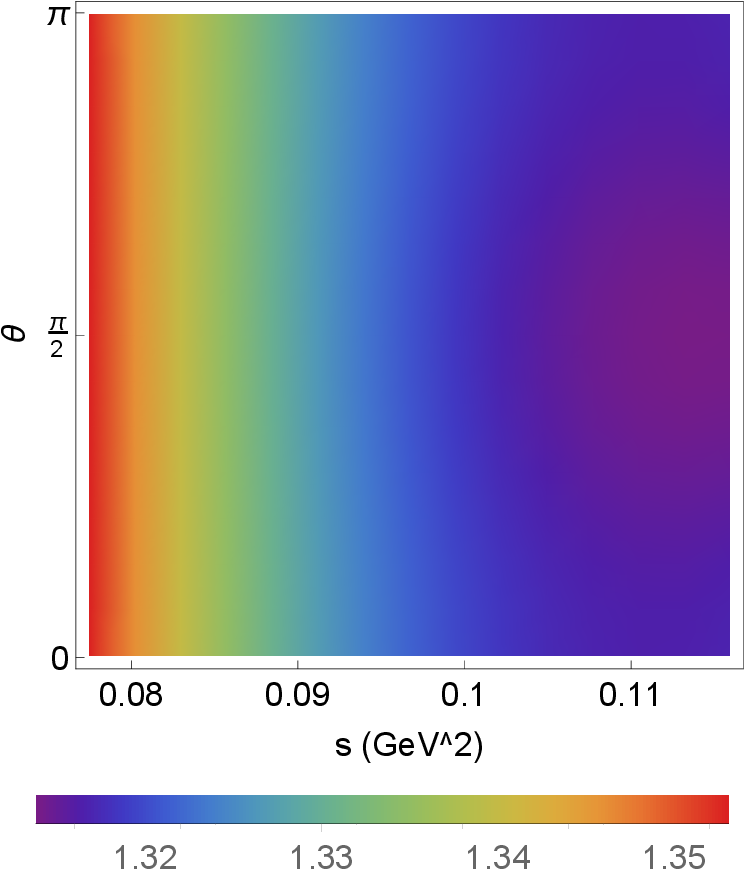}}
\scalebox{0.35}{\includegraphics{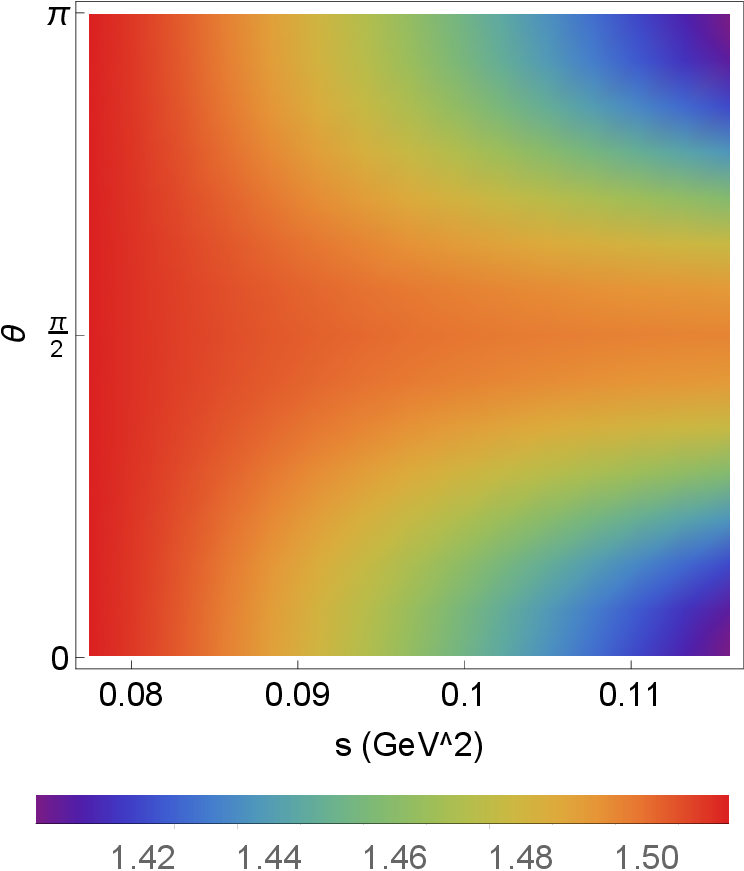}}
\caption{\label{Plot:Entropies_charged}Density plots for the entropies $S_{+0}(s,t,u)$ (left), $S_{00}(s,t,u)$ (center) and $S_{+-}(s,t,u)$ (right) defined in Eq.~\eqref{eq_VNE_charged} as functions of the Mandelstam variable $s$ and the scattering angle $\theta$ in the center-of-mass frame.}
\end{figure}

Figure~\ref{Plot:Entropies_charged} displays density plots of the entropies $S_{+0}$, $S_{00}$, and $S_{+-}$ as functions of the Mandelstam variable $s$ and the center-of-mass scattering angle $\theta$, in the kinematic region defined by\footnote{This kinematic region is chosen to ensure the validity of the one-loop chiral expansion for all the entropies considered; see Sec.~\ref{Sec:Tree-level versus one-loop}. }, $4M_{\pi}^{2}<s<6M_{\pi}^{2}$ and $0<\theta<\pi$. These results confirm and quantify the qualitative discussion presented in Sec.~\ref{qualitative_des_charged}.

After scattering and post-selection, the initially unentangled states $\ket{+0}$, $\ket{0+}$, $\ket{-0}$, and $\ket{0-}$ effectively behave as entangled qubits. Maximal entanglement is reached near threshold and around $\theta=\pi/2$, while the degree of entanglement decreases toward the forward and backward scattering regions as the energy increases.

For the initial states $\ket{00}$, $\ket{+-}$, and $\ket{-+}$, the post-selected final states exhibit qutrit entanglement. Although all three states correspond to three-level systems, their entanglement properties differ quantitatively. The $\ket{00}$ state displays a weak angular dependence and a mild energy variation, whereas the $\ket{+-}$ and $\ket{-+}$ states exhibit a maximum around $\theta=\pi/2$ and a flatter behavior near threshold.  

Figure~\ref{Plot:Entropies_charged} also highlights the different scales over which the entropies $S_{+0}$, $S_{00}$, and $S_{+-}$ vary within the same kinematic region. While $S_{+0}$ exhibits substantial variations, ranging from approximately $0.2$ to $1$, the entropies $S_{00}$ and $S_{+-}$ are considerably more uniform. Across the entire region considered, the latter vary only at the level of a few hundredths, corresponding to relative changes of order $10^{-2}$.

The results of this section demonstrate that, under post-selection, initially separable two-pion flavor states that mix different isospin channels become strongly entangled through strong-interaction dynamics. In this sense, post-selected pion--pion scattering acts as an intrinsic source of flavor entanglement.

This observation naturally raises the complementary question of whether the same mechanism can also suppress or eliminate pre-existing entanglement. This issue is addressed in the following subsection.

\subsection{Reduction of entanglement in post-selected pion scattering}
\label{Sec:reducing_entanglement}

As discussed in Sec.~\ref{sec:pionpionscat}, the isospin channel $I=0$ dominates the pion--pion scattering amplitude over a wide kinematical range. Consequently, any initial two-pion flavor state containing an $I=0$ component is generically driven, after scattering and post-selection, toward a final state dominated by the entangled qutrit $\ket{0,0}$. It follows that entanglement suppression can only occur for initial states with no overlap with the $I=0$ sector. 

Since the only unentangled states in the $\ket{I,M}$ basis are $\ket{2,\pm2}$, a necessary condition for entanglement reduction is that the initial state contains a nonvanishing $\ket{2,\pm2}$ component.

Furthermore, Fig.~\ref{Plot:relative_mag_amplitudes} shows that the $I=1$ amplitude is suppressed relative to the $I=2$ amplitude. This hierarchy allows for the construction of initial superpositions involving $\ket{2,\pm2}$ and $\ket{1,M}$ such that, after scattering and post-selection, the $\ket{2,\pm2}$ sector dominates, leading to a reduction of entanglement in the final state.

Consider, for instance, the initial two-pion state $\ket{s_i}$ defined by
\begin{eqnarray}
    \ket{s_i} = \sqrt{\dfrac{1}{10}} \ket{2,2} + \sqrt{\dfrac{9}{10}} \ket{1,0}, \label{def:initial_entangled_spherical}
\end{eqnarray}
which in the cartesian basis can be rewritten as,
\begin{eqnarray}
    \ket{s_i} = \sum_{cd} U^{11cd} \ket{cd}, \label{def:initial_entangled_spherical}
\end{eqnarray}
where the flavor tensor operator $U^{11cd}$ takes the matrix form
\begin{eqnarray}
    U^{11cd} = \left(
\begin{array}{ccc}
 \frac{1}{2 \sqrt{10}} & -\frac{i \left(\sqrt{2}-6\right)}{4 \sqrt{5}} & 0 \\
 -\frac{i \left(\sqrt{2}+6\right)}{4 \sqrt{5}} & -\frac{1}{2 \sqrt{10}} & 0 \\
 0 & 0 & 0 \\
\end{array}
\right). \label{def:U_initial_entangled}
\end{eqnarray}
For convenience, the first two indices of the flavor tensor $U^{abcd}$ are fixed to $a=b=1$, allowing the reduced flavor density matrix $\rho_{\mathrm{red}}$ to be computed directly using Eq.~\eqref{eq:rhoF_reduced}.

The initial von Neumann entropy $S_i$ associated with the reduced density matrix of the operator $\ket{s_i}\bra{s_i}$, obtained after tracing over the flavor degrees of freedom of one of the particles, is given by
\begin{eqnarray}
    S_i &=& \dfrac{\sqrt{19}-10}{20} \log\left( \dfrac{\sqrt{19}-10}{20} \right) - \dfrac{\sqrt{19}+10}{20} \log\left( \dfrac{\sqrt{19}+10}{20} \right) \nonumber \\
    &\approx& 0.86.  \label{eq:S0}
\end{eqnarray}
This result indicates that the initial state $\ket{s_i}$ is already significantly entangled, with an entropy close to that of a maximally entangled qubit.

Using Eqs.~\eqref{eq:rhoF_reduced} and \eqref{eq:VNentropyeigen}, together with the tensor $U^{abcd}$ defined in Eq.~\eqref{def:U_initial_entangled}, the von Neumann entropy $S_f$ of the state $\ket{s_i}$ after scattering and post-selection can be written as
\begin{eqnarray}
    S_f = - \sum_{i=1}^2 \lambda_i \log(\lambda_i), \label{eq:entropy_entangled_initial_after}
\end{eqnarray}
where the eigenvalues $\lambda_i$ are
\begin{eqnarray}
    \lambda_1 &=& \dfrac{1}{20 N_i} \left( 9 |A_1|^2 + |A_2|^2 - |A_2| \sqrt{18 |A_1|^2+|A_2|^2} \right), \nonumber \\
    \lambda_2 &=& \dfrac{1}{20 N_i} \left( 9 |A_1|^2 + |A_2|^2 + |A_2| \sqrt{18 |A_1|^2+|A_2|^2} \right), \nonumber \\
    N_i &=& \dfrac{9}{10} |A_1|^2 + \dfrac{1}{10} |A_2|^2, \label{def:S_lambdas_initial_entangled}
\end{eqnarray}
with the isospin amplitudes $A_1$ and $A_2$ defined in Eq.~\eqref{def:ATotalIsospin}.

\begin{figure}[h]
\scalebox{0.35}{\includegraphics{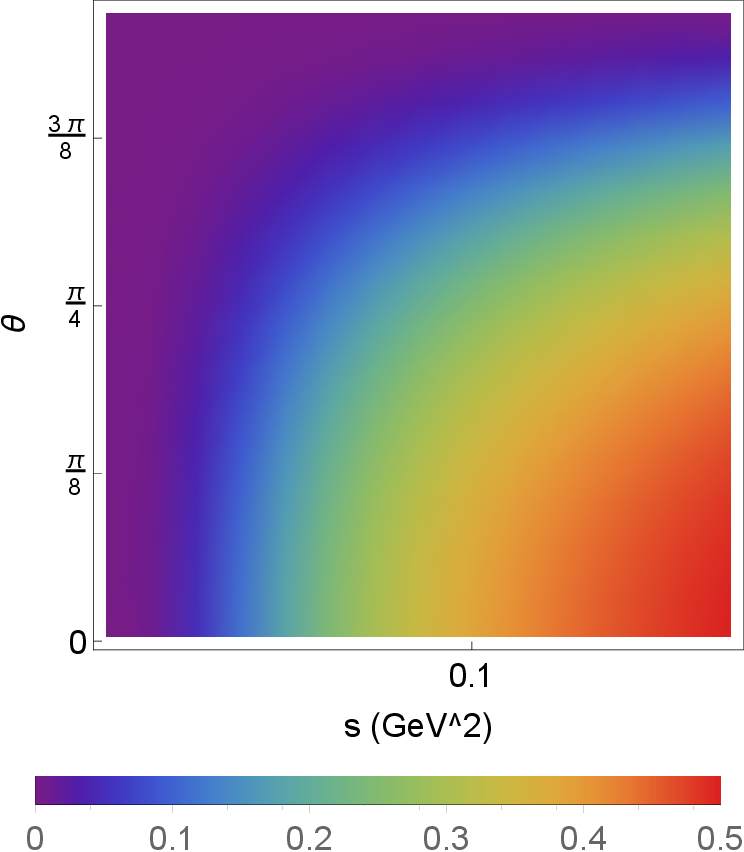}}
\scalebox{0.35}{\includegraphics{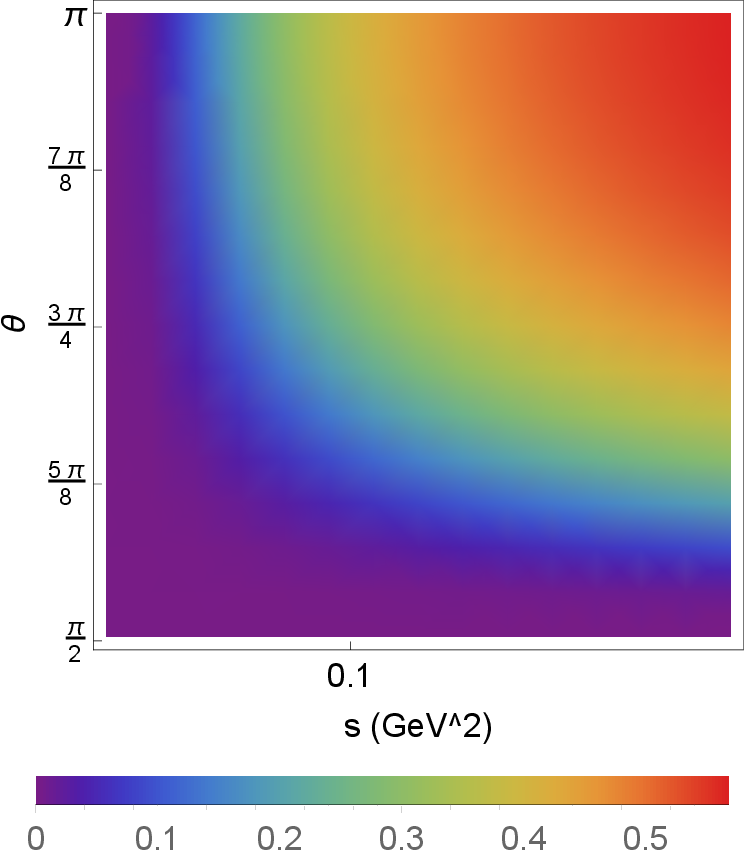}}
\caption{\label{Plot:Entropies_initial_entangled}
Density plots of the entropy $S_{f}(s,t,u)$, defined in Eq.~\eqref{eq:entropy_entangled_initial_after}, shown as a function of the Mandelstam variable $s$ and the scattering angle $\theta$ in the center-of-mass frame. The left panel corresponds to the kinematical region $0 \leq \theta \leq \pi/2$, while the right panel shows $\pi/2 \leq \theta \leq \pi$.
}
\end{figure}

At threshold, the entropy $S_f$ in Eq. \eqref{def:S_lambdas_initial_entangled} simplifies to
\begin{eqnarray}
    S_f = 0.
    \label{eq:S_f_threshold}
\end{eqnarray}
As expected, Eq.~\eqref{eq:S_f_threshold} shows that, at threshold, the initially entangled state $\ket{s_i}$ becomes completely disentangled after scattering and post-selection. In fact, near threshold, the entropy $S_f$ remains close to zero. As the energy increases up to $s=0.1\, \text{Gev}^2$, the entropy begins to increase as well, as illustrated in Fig. \ref{Plot:Entropies_initial_entangled}. For clarity, the figure is split into two panels, since at $\theta = \frac{\pi}{2}$ the entropy develops logarithmic singularities of the form $0 \log 0$.

More generally, any normalized entangled initial state of the form
\begin{eqnarray}
\ket{s_i} = \alpha_{2,\pm 2}\ket{2,\pm 2} + \sum_M \alpha_{1,M}\ket{1,M},
\label{General_suppressed_EntSate}
\end{eqnarray}
experiences a reduction of its entanglement following scattering and post-selection near threshold. This behavior arise from the vanishing of the total-isospin amplitude $A_1(s,t,u)$, defined in Eq.~\ref{def:ATotalIsospin}, at threshold at one-loop order. As a result, the $I=1$ components of the initial state are dynamically suppressed and are not propagated into the final scattering state in this kinematic regime. The post-selected final state is therefore projected onto the pure unentangled state $\ket{2,\pm 2}$, leading to the complete suppression of the initial entanglement. This behavior is analogous to that identified in Ref.~\cite{Blasone:2024_2} for tree-level QED scattering processes. 

\subsection{Tree-level versus one-loop contributions}
\label{Sec:Tree-level versus one-loop}

\begin{figure}[h]
\scalebox{0.35}{\includegraphics{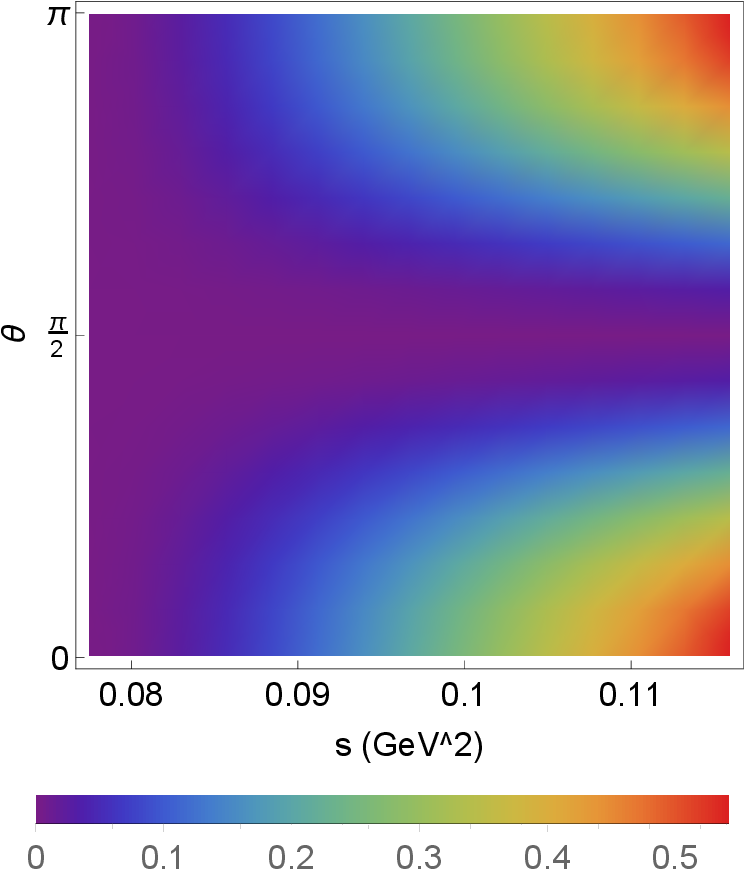}}
\scalebox{0.35}{\includegraphics{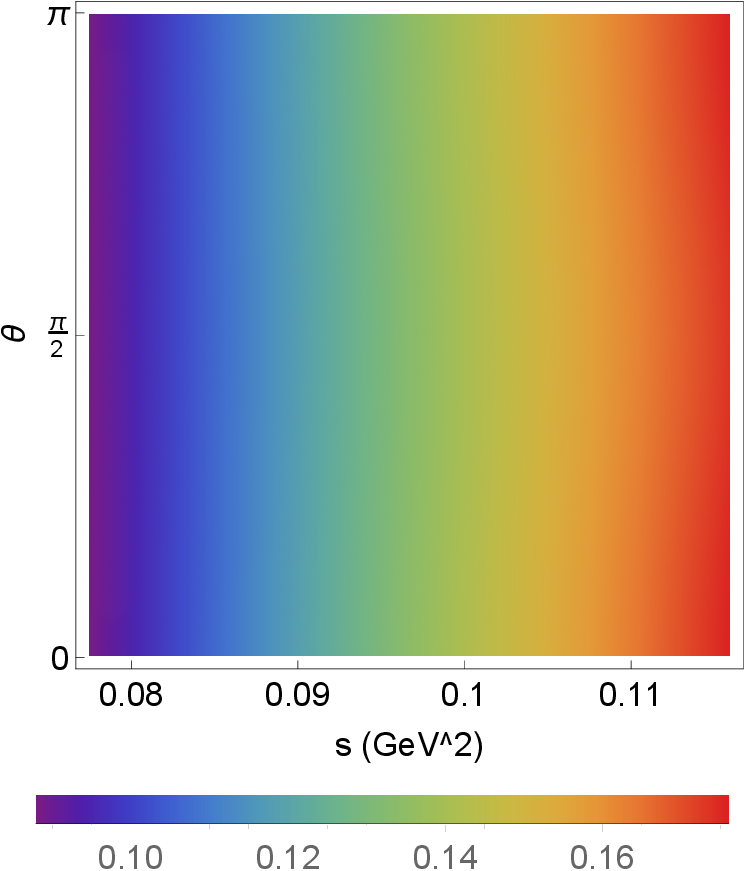}}
\scalebox{0.35}{\includegraphics{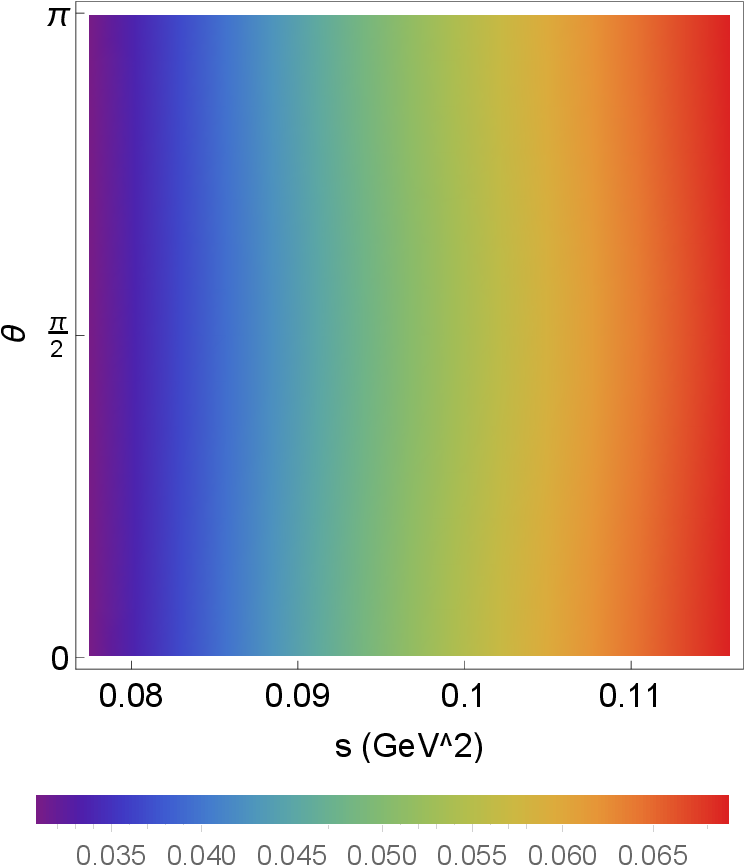}}
\caption{\label{Plot:Diff_Entropies_charged}Density plots of the relative difference $|S_{\text{1-Loop}}-S_{\text{Tree}}|/S_{\text{Tree}}$ for the entropies associated with the $+0$ (left), $00$ (center), and $+-$ (right) channels, as defined in Eq.~\eqref{eq_VNE_charged}, shown as functions of the Mandelstam variable $s$ and the center-of-mass scattering angle $\theta$.}
\end{figure}

\begin{figure}[h]
\scalebox{0.35}{\includegraphics{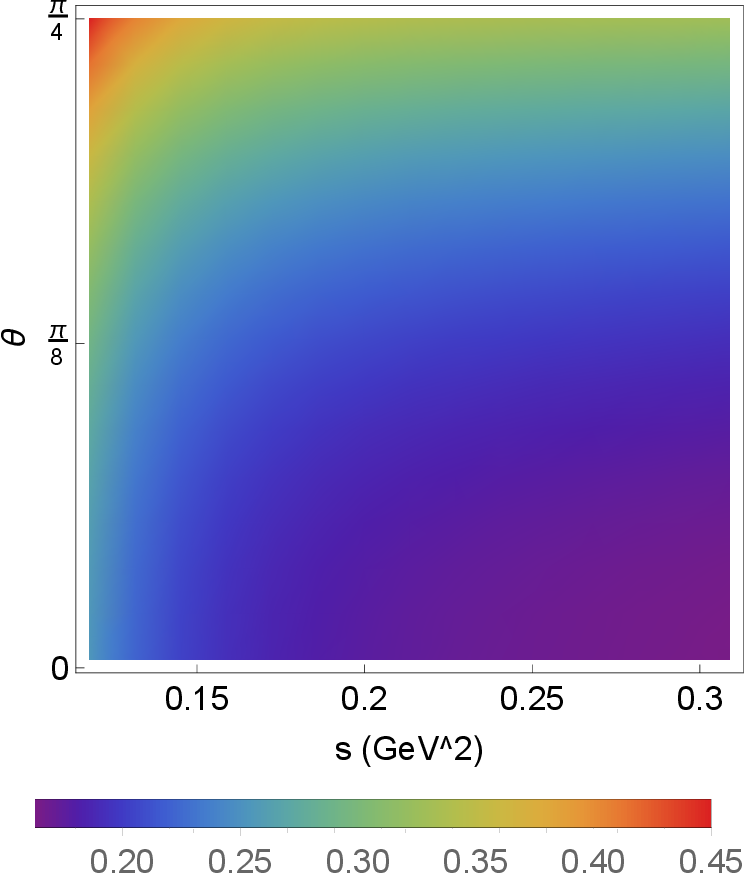}}
\scalebox{0.35}{\includegraphics{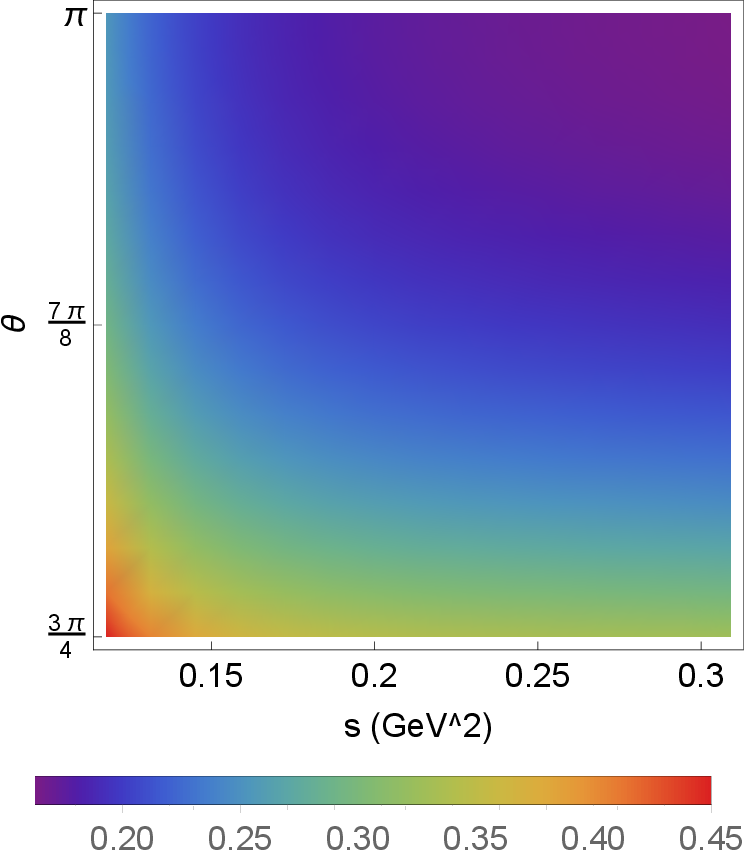}}
\caption{\label{Plot:Diff_Entropies_initial_entangled}
Density plots of the relative difference $(S_{\text{1-Loop}}-S_{\text{Tree}})/S_{\text{Tree}}$ for the entropy defined in Eq.~\eqref{eq:entropy_entangled_initial_after}, shown as a function of the Mandelstam variable $s$ and the center-of-mass scattering angle $\theta$. The displayed kinematic regions exclude the suppression zones where the entropy vanishes or becomes negligibly small.}
\end{figure}

At tree level, the scattering amplitude $A(s,t,u)$ defined in
Eq.~\eqref{eq:Tmtxelem} reduces to the simple form
\begin{equation}
A(s,t,u)=\frac{s-M_\pi^2}{F_\pi^2}.
\label{eq:ATreePhysical}
\end{equation}
Substituting this expression into the von Neumann entropies defined in Eqs.~\eqref{eq_VNE_charged} and \eqref{eq:entropy_entangled_initial_after},
and comparing the resulting tree-level entropies with their one-loop counterparts, the density plots shown in Figs.~\ref{Plot:Diff_Entropies_charged} and \ref{Plot:Diff_Entropies_initial_entangled} are obtained. 

The kinematic region considered in Fig.~\ref{Plot:Diff_Entropies_charged} is chosen to ensure the validity of the perturbative expansion. Throughout most of this region, one-loop corrections modify the tree-level value of $S_{+0}$ by less than $50\%$. As the center-of-mass energy increases, however, the relative correction becomes larger in the corners of the phase space corresponding to forward and backward scattering. In these kinematic configurations, either $t$ or $u$ approaches zero, causing the one-loop contributions to $A_t(s,t,u)$ and $A_u(s,t,u)$ to become comparable to their respective tree-level values as $s$ increases. Since the entropy $S_{+0}$ depends exclusively on these amplitudes, its perturbative domain of validity is more restricted than those of $S_{00}$ and $S_{+-}$, which remain perturbatively stable over a broader kinematic region.

For the initially unentangled charged states, the tree-level amplitudes already generate nontrivial flavor entanglement after scattering and post-selection. This entanglement originates from the interference among different isospin channels and does not rely on loop effects. The inclusion of one-loop corrections does not qualitatively modify this mechanism. Instead, loop contributions induce controlled, channel-dependent quantitative changes in the entropies.

In particular, for the $+0$ channel, one-loop effects reduce\footnote{The relative difference $(S_{\text{1-Loop}}-S_{\text{Tree}})/S_{\text{Tree}}$ is negative for $S_{+0}$ and positive for $S_{00}$ and $S_{+-}$ throughout the kinematic region shown in Fig.~\ref{Plot:Diff_Entropies_initial_entangled}.} the tree-level entropy, resulting in a sharper entanglement peak near threshold and around the central scattering angle $\theta=\pi/2$. By contrast, for the $00$ and $+-$ channels, loop corrections systematically enhance the entropy throughout the phase space. In the $00$ channel, these corrections distort the band-like structures present at tree level, while in the $+-$ channel they strengthen the angular dependence of the entropy, leading to a more pronounced maximum around $\theta=\pi/2$. Overall, these results demonstrate that quantum corrections not only modify the magnitude of the entanglement entropy but also refine its kinematic structure relative to the tree-level prediction.

For initially entangled states, one-loop corrections increase the entropy relative to its tree-level value, thereby enhancing entanglement generation outside the regions of entanglement suppression located near threshold and around the central scattering angle $\theta=\pi/2$ (see Fig.~\ref{Plot:Entropies_initial_entangled}). These results suggest that quantum fluctuations do more than simply rescale the tree-level predictions; they also redistribute entanglement across the available kinematic configurations.

\section{\label{sec:Conclusions}Conclusions}

This work has investigated the generation and suppression of flavor entanglement in pion-pion scattering within a post-selected measurement framework. In contrast to approaches based on the unitary $S$-matrix description of scattering processes, the present analysis conditions the final state on a post-selection of outgoing momenta. This allows the entanglement associated with specific scattering events to be characterized. The post-selection procedure with fixed momenta is formulated in Sec.~\ref{sec:Postselected_formalism} using a wave-packet description.

The amount of flavor entanglement generated after scattering and post-selection is quantified using the quantum mutual information of the pion flavor degrees of freedom. This quantity requires the evaluation of the von Neumann entropy of the reduced flavor density matrix $\rho_{\mathrm{red}}$, whose general expression for arbitrary initial states is derived in Eq.~\eqref{eq:rhoF_reduced}. The reduced density matrix depends on the initial configuration through the flavor tensor operator $U^{abcd}$ defined in Eq.~\eqref{eq:initalstate}, and the corresponding entanglement entropy follows directly from its eigenvalues.

Within this post-selected formalism, the flavor entanglement generated from initially unentangled charged two-pion states is analyzed in Sec.~\ref{Sec:entropy_chargedstates}. Since the strong interaction conserves total isospin, the states $\ket{++}$ and $\ket{--}$ are not modified by the scattering process and therefore remain unentangled. In contrast, the remaining charged states develop nontrivial entanglement and can be classified, after scattering and post-selection, as effective qubit or qutrit systems depending on their isospin content.

Initial charged states containing an $I=0$ component, namely $\ket{+-}$, $\ket{-+}$, and $\ket{00}$, evolve into entangled qutrit states. Their entanglement entropies at threshold are given in Eq.~\eqref{eq:VN_charged_threshold_numerical}, while their dependence on scattering angle and energy is illustrated in Fig.~\ref{Plot:Entropies_charged}. Although these states do not reach the maximal qutrit entropy $\log(3)$, their entropies remain close to this value across the kinematical region considered and lie between those of the spherical states $\ket{0,0}$ and $\ket{2,0}$ (see Table~\ref{table:SphericalBasisVNE}).

The remaining charged states $\ket{+0}$, $\ket{0+}$, $\ket{-0}$, and $\ket{0-}$ behave as effective qubits after scattering and post-selection. At threshold, their entanglement entropy reaches the maximal qubit value $S=1$ at order $\mathcal{O}(1/F_\pi^2)$, independently of the low-energy constants. In addition, the entropies associated with the states $\ket{+-}$, $\ket{-+}$ and $\ket{\pm 0}$, $\ket{0 \pm}$ exhibit a common angular dependence, with a maximum near the scattering angle $\theta = \pi/2$. Qualitatively similar angular behavior has been reported in studies of entanglement generation in QED scattering processes \cite{Cervera-Lierta,Blasone:2024}.

The dominant role of the $I=0$ isospin channel in pion-pion scattering, illustrated in Fig.~\ref{Plot:relative_mag_amplitudes}, implies that this channel provides the leading contribution to the scattering amplitude. As a consequence, after scattering and post-selection with fixed momenta, initial states containing a $\ket{0,0}$ component are preferentially projected onto entangled qutrit configurations.

The analysis further shows that the pion-pion interaction in the post-selection framework can also lead to a reduction of entanglement under suitable conditions. As demonstrated in Sec.~\ref{Sec:reducing_entanglement}, initial states constructed as superpositions of $\ket{2,\pm 2}$ and $I=1$ components may lose their initial entanglement after scattering and post-selection. This behavior can be traced to the relative enhancement of the $I=2$ scattering amplitude compared with the $I=1$ channel, together with the fact that the states $\ket{2,\pm 2}$ are intrinsically unentangled.

Finally, a comparison between tree-level and one-loop results shows that tree-level amplitudes already capture the qualitative entanglement patterns discussed in Secs.~\ref{Sec:entropy_chargedstates} and \ref{Sec:reducing_entanglement}. However, one-loop corrections do not simply rescale these results, but instead introduce systematic, channel-dependent modifications. In particular, they redistribute entanglement across the pion-pion phase space, enhancing specific kinematical regions and sharpening angular structures that appear more diffuse at tree level.

The mechanisms identified in this work rely on symmetry properties, channel dominance, and post-selection, and are therefore not specific to pion-pion scattering. Similar entanglement generation and suppression effects may arise in other non-Abelian scattering processes, such as kaon, nucleon, or heavy-meson interactions.

Within the post-selection framework adopted in this work, an important open question concerns the role of the momentum-space wave-packet width $\sigma$ defined in Eq. \ref{eq:wavepacket}. The present analysis has focused on the limit $\sigma\to 0$, whereas finite wave-packet widths generate corrections of order $\mathcal{O}(\sigma^2)$ to the von Neumann entropies. Even for relatively small values of $\sigma$, such contributions may become comparable to one-loop effects in certain kinematic regimes and could therefore influence both the generation and suppression of entanglement. Understanding this interplay represents a natural next step toward connecting the present formalism with realistic scattering experiments and lattice-inspired implementations.

\section*{Acknowledgments}
The authors are grateful to {\em Secretaría de Ciencia, Humanidades, Tecnología e Innovación} SECIHTI (Mexico) for support through the {\it Ciencia de Frontera} project CF-2023-I-162.

\appendix

\section{\label{App:ScatteringAmplitudeOneloop}Pion--pion scattering at one-loop order}

This appendix summarizes the derivation of the pion--pion scattering amplitude $A(s,t,u)$ defined in Eq.~\eqref{eq:Tmtxelem}, obtained from the quantum effective action $\Gamma[\phi]$ within chiral perturbation theory. The presentation is included for completeness and to make explicit the steps leading to the one-loop amplitude employed in the main text.

At one-loop order and in $D$ spacetime dimensions, the quantum effective action is given by \cite{peskin}
\begin{equation}
\Gamma[\phi]
=
\int d^Dx\, \mathcal{L}
+\frac{i}{2}\,
\ln\!\left[
\mathrm{Det}\!\left(
-\frac{\delta}{\delta \phi^b(y)\,\delta\phi^a(x)}
\int d^Dz\,\mathcal{L}_2
\right)
\right],
\label{eq:defQAoneloop_app}
\end{equation}
where the functional derivative satisfies
\begin{equation}
\frac{\delta}{\delta \phi^b(y)}\,\phi^a(x)
=
\delta^{ab}\,\delta(x-y),
\label{functional_derv_def_app}
\end{equation}
and the Lagrangians $\mathcal{L}$ and $\mathcal{L}_2$ are defined in Eqs.~\eqref{eq:ChPTLag} and \eqref{eq:Lag2}, respectively.

The operators $L_i$ and $H_i$ entering the $\mathcal{O}(p^4)$ Lagrangian $\mathcal{L}_4$, defined in Eq.~\eqref{eq:Lag4}, are listed explicitly as
\begin{eqnarray}
L_1 &=& \frac{1}{4}
\left\{
\mathrm{Tr}\!\left[
D_\mu U (D^\mu U)^\dagger
\right]
\right\}^2,
\nonumber \\
L_2 &=& \frac{1}{4}
\mathrm{Tr}\!\left[D_\mu U(D_\nu U)^\dagger\right]
\mathrm{Tr}\!\left[D^\mu U(D^\nu U)^\dagger\right],
\nonumber \\
L_3 &=& \frac{1}{16}
\left[
\mathrm{Tr}(\chi U^\dagger + U \chi^\dagger)
\right]^2,
\nonumber \\
L_4 &=& \frac{1}{4}
\mathrm{Tr}\!\left[
D_\mu U(D^\mu \chi)^\dagger
+
D_\mu \chi (D^\mu U)^\dagger
\right],
\nonumber \\
L_5 &=&
\mathrm{Tr}(f_{R\mu\nu} U f_L^{\mu\nu} U^\dagger)
-\frac12 \mathrm{Tr}(f_{L\mu\nu}f_L^{\mu\nu}+f_{R\mu\nu}f_R^{\mu\nu}),
\nonumber \\
L_6 &=& \frac{i}{2}
\mathrm{Tr}\!\left[
f_{R\mu\nu}D^\mu U(D^\nu U)^\dagger
+
f_{L\mu\nu}(D^\mu U)^\dagger D^\nu U
\right],
\nonumber \\
L_7 &=&
-\frac{1}{16}
\left[
\mathrm{Tr}(\chi U^\dagger-U\chi^\dagger)
\right]^2 .
\label{def:L_operators_L4_app}
\end{eqnarray}

The corresponding $H_i$ operators read
\begin{eqnarray}
H_1 &=&
\frac{1}{16}
\Big\{
[\mathrm{Tr}(\chi U^\dagger+U\chi^\dagger)]^2
+
[\mathrm{Tr}(\chi U^\dagger-U\chi^\dagger)]^2
\nonumber\\
&&\hspace{0.5cm}
-2\,\mathrm{Tr}(\chi U^\dagger \chi U^\dagger + U\chi^\dagger U\chi^\dagger)
\Big\}
+\frac14 \mathrm{Tr}(\chi\chi^\dagger),
\nonumber\\
H_2 &=&
-2\,\mathrm{Tr}(f_{L\mu\nu}f_L^{\mu\nu}+f_{R\mu\nu}f_R^{\mu\nu}),
\nonumber\\
H_3 &=&
-\frac{1}{16}
\Big\{
[\mathrm{Tr}(\chi U^\dagger+U\chi^\dagger)]^2
+
[\mathrm{Tr}(\chi U^\dagger-U\chi^\dagger)]^2
\nonumber\\
&&\hspace{0.5cm}
-2\,\mathrm{Tr}(\chi U^\dagger \chi U^\dagger + U\chi^\dagger U\chi^\dagger)
\Big\}
+\frac14 \mathrm{Tr}(\chi\chi^\dagger).
\label{def:H_operators_L4_app}
\end{eqnarray}

The field-strength tensors $f^L_{\mu\nu}$ and $f^R_{\mu\nu}$ are defined in terms of the external sources $l_\mu$ and $r_\mu$ as
\begin{subequations}
\begin{eqnarray}
f^L_{\mu\nu} &=& \partial_\mu l_\nu - \partial_\nu l_\mu - i[l_\mu,l_\nu],\\
f^R_{\mu\nu} &=& \partial_\mu r_\nu - \partial_\nu r_\mu - i[r_\mu,r_\nu],
\end{eqnarray}
\end{subequations}
with
\begin{subequations}
\begin{eqnarray}
l_\mu &=& \tau^a l^a_\mu,\\
r_\mu &=& \tau^a r^a_\mu,
\end{eqnarray}
\end{subequations}
where $l^a_\mu$ and $r^a_\mu$ are real external fields.

The covariant derivative acting on $U$ takes the form
\begin{equation}
D_\mu U = \partial_\mu U + i l_\mu U - i r_\mu U.
\label{def_covariant_der_app}
\end{equation}

The scalar and pseudoscalar sources are combined into
\begin{equation}
\chi(x)=2B_0\,[s(x)+ip(x)],
\end{equation}
with
\begin{eqnarray}
s(x) &=& s^0(x)+\tau^a s^a(x),\\
p(x) &=& p^0(x)+\tau^a p^a(x),
\end{eqnarray}
where all components are real.

To evaluate the $T$-matrix element in Eq.~\eqref{eq:Tmtxelem}, the external fields are set to zero, $l_\mu=r_\mu=0$, and the isospin-symmetric limit $\chi=2B_0 m\,\mathds{1}$ is adopted. The $T$-matrix element is then obtained from the effective action via the LSZ reduction formula,
\begin{equation}
T^{abcd}
=
\mathcal{R}^2\,
\Gamma^{abcd}(p_1,p_2,p_3,p_4),
\label{T_matrix_Gamma_app}
\end{equation}
where $\mathcal{R}$ denotes the residue of the exact pion propagator at $p^2=M_\pi^2$.

The amputated four-point function $\Gamma^{abcd}$ is defined as
\begin{eqnarray}
(2\pi)^D \delta\!\left(\sum_{i=1}^4 p_i\right)
\Gamma^{abcd}
&=&
\int \prod_{i=1}^4 d^Dx_i\,
e^{i\sum_i p_i\cdot x_i}
\nonumber\\
&&\times
\left.
\frac{\delta^4 \Gamma[\phi]}
{\delta\phi^d(x_4)\delta\phi^c(x_3)\delta\phi^b(x_2)\delta\phi^a(x_1)}
\right|_{\phi=0}.
\label{eq:FTGammascat_app}
\end{eqnarray}

The Lagrangians $\mathcal{L}_2$ and $\mathcal{L}_4$ are expanded up to $\mathcal{O}(\phi^6)$ and $\mathcal{O}(\phi^4)$, respectively, which is sufficient for the one-loop four-point function,
\begin{subequations}
\begin{eqnarray}
\mathcal{L}_2 &=& \mathcal{L}_2^{2\phi}+\mathcal{L}_2^{4\phi}+\mathcal{L}_2^{6\phi},\\
\mathcal{L}_4 &=& \mathcal{L}_4^{2\phi}+\mathcal{L}_4^{4\phi}.
\end{eqnarray}
\end{subequations}
where
\begin{subequations}
\label{eq:Lcontorderphi}
\begin{eqnarray}
\mathcal{L}_2^{2\phi} & = & \frac{1}{2} \left( \partial_\mu \bm{\phi} \cdot \partial^\mu \bm{\phi} - M^2 \bm{\phi} \cdot \bm{\phi} \right), \\
\mathcal{L}_2^{4\phi} & = & \frac{M^2}{24F^2} (\bm{\phi}\cdot \bm{\phi})^2 + \frac{1}{6 F^2} \left( \bph \cdot \partial_\mu \bph \, \bph \cdot \partial^\mu \bph - \bph \cdot \bph \, \partial_\mu \bph \cdot \partial^\mu \bph \right), \\
\mathcal{L}_2^{6\phi} & = & \frac{1}{45 F^4} \partial_\mu \bph \cdot \partial^\mu \bph \, (\bph \cdot \bph)^2 - \frac{1}{45 F^4} \partial_\mu \bph \cdot \bph \, \partial^\mu \bph \cdot \bph \, \bph \cdot \bph \nonumber \\
&& - \frac{M^2}{720 F^4} (\bph \cdot \bph)^3, \\
\mathcal{L}_4^{2\phi} & = & - \frac{l_3 M^4}{F^2} \bph \cdot \bph, \\
\mathcal{L}_4^{4\phi} & = & \frac{l_1}{F^4} \left( \partial_\mu \bph \cdot \partial^\mu \bph \right)^2 + \frac{l_2}{F^4} \partial_\mu \bph \cdot \partial_\nu \bph \, \partial^\mu \bph \cdot \partial^\nu \bph + \frac{l_3 M^4}{3F^4} (\bph \cdot \bph)^2.
\end{eqnarray}
\end{subequations}
Here, $\bm{\phi} = (\phi^1, \phi^2, \phi^3)$, the mass parameter is defined as usual, $M = 2 B_0 m$, and terms independent of $\phi$ have been omitted. 

The logarithmic determinant in Eq.~\eqref{eq:defQAoneloop_app} is expanded using $\mathrm{Det}(A)=\e^{\mathrm{Tr}(\ln A)}$, yielding
\begin{eqnarray}
\ln \mathrm{Det}(\cdots)
&=& \mathrm{Tr}\left(\ln\left[ \Delta^{a b}(x-y) \right]\right) \nonumber \\
& & \mbox{\hglue-1.6truecm} - \sum_{n=1}^\infty \frac{1}{n} \int d^Dz_1 \dots d^Dz_n \Delta^{a_1 a_2}(z_1-z_2) \left[ \frac{\delta }{\delta \phi^{a_2}(z_2)\delta\phi^{a_3}(z_3)} \int d^Dz\, (\mathcal{L}_2^{4\phi}+\mathcal{L}_2^{6\phi}) \right] \nonumber \\
& & \mbox{\hglue-1.6truecm} \times \Delta^{a_3 a_4}(z_3-z_4) \dots \left[ \frac{\delta }{\delta \phi^{a_{2n}}(z_{2n})\delta\phi^{a_1}(z_1)} \int d^Dz\, (\mathcal{L}_2^{4\phi}+\mathcal{L}_2^{6\phi}) \right],
\label{eq:Logfunct_app}
\end{eqnarray}
where
\begin{equation}
\Delta^{ab}(x-y)
=
-\delta^{ab}\int\!\frac{d^Dk}{(2\pi)^D}
\frac{e^{ik\cdot(x-y)}}{k^2-M^2}. \label{Prop_xspace}
\end{equation}

Thus, Eq.~\eqref{eq:FTGammascat_app} for $\Gamma^{ab cd} (p_1,p_2,p_3,p_4)$ at one-loop order becomes,
\begin{eqnarray}
&&(2\pi)^D \delta\left(\sum_{i=1}^4 p_i\right)\Gamma^{ab cd}(p_1,p_2,p_3,p_4) = \nonumber \\ 
&& \hspace{10pt} \int \left(\prod_{i=1}^4 d^Dx_i\right) \e^{i \sum_{i=1}^4 p_i \cdot x_i} \, \frac{\delta}{\delta \phi^d(x_4)\delta \phi^c(x_3)\delta \phi^b(x_2)\delta \phi^a(x_1)} \left( \sum_{i=1}^4 \bar{\Gamma}_i \right), \label{eq:FTGammascat_second_form}
\end{eqnarray}
where
\begin{subequations}
\label{eq:defbarGam}
\begin{eqnarray}
\bar{\Gamma}_1 & = & \int d^Dx \, \mathcal{L}_2^{4\phi}, \\
\bar{\Gamma}_2 & = & \int d^Dx \, \mathcal{L}_4^{4\phi}, \\
\bar{\Gamma}_3 & = & - \frac{i}{2} \int d^Dz_1 d^Dz_2 \Delta^{a_1a_2}(z_1-z_2) \frac{\delta }{\delta \phi^{a_2}(z_2)\delta\phi^{a_1}(z_1)} \int d^Dz\, \mathcal{L}_2^{6\phi}, \\
\bar{\Gamma}_4 & = & - \frac{i}{4} \int \left(\prod_{i=1}^4d^Dz_i\right) \Delta^{a_1 a_2}(z_1-z_2) \left[ \frac{\delta }{\delta \phi^{a_2}(z_2)\delta\phi^{a_3}(z_3)} \int d^D z\, \mathcal{L}_2^{4\phi} \right]   \nonumber \\
& & \mbox{} \times \Delta^{a_3 a_4}(z_3-z_4) \left[ \frac{\delta }{\delta \phi^{a_4}(z_4)\delta\phi^{a_1}(z_1)} \int d^Dz^\prime\, \mathcal{L}_2^{4\phi} \right].
\end{eqnarray}
\end{subequations}

The amplitude $A(s,t,u)$ defined in Eq.~\eqref{eq:Tmtxelem} receives contributions from the four distinct terms $\bar{\Gamma}_i$, leading to
\begin{equation}
A(s,t,u)=\mathcal{R}^2\sum_{i=1}^4 A_i(s,t,u).
\end{equation}

By applying the functional derivatives directly to the Lagrangian term
$\mathcal{L}_2^{4\phi}$ and using Eq.~\eqref{Gamma_4phi}, the contribution
$A_1(s,t,u)$ at order $\mathcal{O}(F^{-4})$ is obtained as
\begin{equation}
A_1(s,t,u) = \frac{s-M^2}{F^2} + \frac{M^4}{24\pi^2F^4} \overline{l}_3, \label{A1}
\end{equation}
where the result for the one-loop correction to the pion mass, given in Eq.~\eqref{eq:masspiononeloop}, has been used.

Similarly, the contribution $A_2(s,t,u)$ arising from $\bar{\Gamma}_2$ is found to be
\begin{equation}
A_2(s,t,u) = \frac{8M^4}{3F^4}l_3 - \frac{l_2}{F^4}\left( 8 M^4-4M^2s-t^2-u^2 \right) + \frac{2l_1}{F^4}(s-2M^2)^2. \label{A2}
\end{equation}
This expression follows from imposing the on-shell condition
$p_i^2 = M^2$, which is valid at order $\mathcal{O}(F^{0})$.
The loop correction to the pion mass in Eq.~\eqref{eq:masspiononeloop}
is not included here, as it would generate terms of order
$\mathcal{O}(F^{-6})$, beyond the accuracy of the present calculation.

The contribution $A_3(s,t,u)$ originates from the sixth-order interaction term $\mathcal{L}_2^{6\phi}$ and requires the evaluation of the corresponding sixth-order functional derivative contracted with a single propagator. Proceeding analogously to the previous cases, a straightforward calculation yields,
\begin{equation}
A_3(s,t,u) = \frac{1}{18F^4}(31M^2-20s) J_1, \label{A3}
\end{equation}
where the scalar Feynman integral $J_1$ is defined in
Eq.~\eqref{Iden_tadpole_int}, and the leading-order on-shell approximation $p_i^2 = M^2$ has again been employed.

The evaluation of the contribution $A_4(s,t,u)$, which arises from $\bar{\Gamma}_4$, is facilitated by the identity
\begin{eqnarray}
& & \int \left(\prod_{i=1}^4 d^Dx_i\right) \left(\prod_{i=1}^4 d^Dz_i\right) \e^{i \sum_{i=1}^4 p_i \cdot x_i} \Delta^{a_1 a_2}(z_1-z_2) \Delta^{a_3 a_4}(z_3-z_4) F^{abcda_1a_2a_3a_4} \nonumber \\
& & \mbox{\hglue0.2truecm} = (2\pi)^D \delta\left(\sum_{i=1}^4p_i\right) \int \frac{d^Dk}{(2\pi)^D} \left( \frac{1}{(k^2-M^2)[(p_1+p_2+k)^2-M^2]} \right) \nonumber \\
& & \mbox{\hglue0.6truecm} \times \Lambda^{a_1a_2ab}(k,-p_1-p_2-k,p_1,p_2) \Lambda^{a_2a_1cd}(p_1+p_2+k,-k,p_3,p_4), \label{eq:identforA4}
\end{eqnarray}
where the function $\Lambda^{abcd}$ is given by Eq.~\eqref{def_Lambda} and $F^{abcda_1a_2a_3a_4}$ is defined as
\begin{eqnarray}
F^{abcda_1a_2a_3a_4}  &=&  \left( \frac{\delta}{\delta \phi^d(x_4)\delta \phi^c(x_3)\delta \phi^{a_2}(z_2)\delta \phi^{a_3}(z_3)} \int d^Dz \mathcal{L}_2^{4\phi} \right) \nonumber \\
& &  \times  \left( \frac{\delta}{\delta \phi^b(x_2)\delta \phi^a(x_1)\delta \phi^{a_4}(z_4)\delta \phi^{a_1}(z_1)} \int d^Dz^\prime \mathcal{L}_2^{4\phi} \right). \label{def_F}
\end{eqnarray}

The product of $\Lambda$ tensors appearing in Eq.~\eqref{eq:identforA4} can be rewritten as
\begin{eqnarray}
& & \Lambda^{a_1a_2ab}(k,-p_1-p_2-k,p_1,p_2) \Lambda^{a_2a_1cd}(p_1+p_2+k,-k,p_3,p_4)  = \nonumber \\
&& \mbox{\hglue1.0truecm} \frac{1}{9F^4} \Bigg\{ \left[3M^2\left( \Lambda_1^{(3)}  + \Lambda_2^{(3)} \right) + \Lambda_1^{(3)} \Lambda_2^{(3)} \right] \delta^{ab}\delta^{cd}  + \left[\Lambda_1^{(1)}\Lambda_2^{(2)}+\Lambda_1^{(2)}\Lambda_2^{(1)}\right]\delta^{ac}\delta^{bd} \nonumber \\
& & \mbox{\hglue1.0truecm} + \left[ \Lambda_1^{(1)} \Lambda_2^{(1)} + \Lambda_1^{(2)} \Lambda_2^{(2)} \right] \delta^{ad} \delta^{bc} \Bigg\}, \label{prod_Lambdas}
\end{eqnarray}
where
\begin{subequations}
\label{Lambdas12}
\begin{eqnarray}
\Lambda_1^{(i)} & = & \Lambda^{(i)}(k,-p_1-p_2-k,p_1,p_2), \\
\Lambda_2^{(i)} & = & \Lambda^{(i)}(p_1+p_2+k,-k,p_3,p_4),
\end{eqnarray}
\end{subequations}
with the functions $\Lambda^{(i)}$ defined in Eq.~\eqref{Lambdas}.

Combining Eqs.~\eqref{eq:identforA4} and \eqref{Lambdas12} with the results summarized in Appendix~\ref{app:feynmanint}, the contribution $A_4(s,t,u)$ is finally obtained as,
\begin{eqnarray}
A_4(s,t,u) & = & \frac{1}{18 (D-1) F^4} \Big[ (22+5D)s - 4(8+D)M^2 \Big] J_1 - \frac{1}{2 F^4}(s^2-M^4) J_2(s) \nonumber \\
& & \hspace{-50pt} - \frac{1}{4F^4(D-1)} \left[4(D+3)M^4-4M^2(Dt+2s+t)+t(Dt+2s)\right] J_2(t) \nonumber \\
& & \hspace{-50pt} - \frac{1}{4F^4(D-1)} \left[4(D+3)M^4-4M^2(Du+2s+u)+u(Du+2s)\right] J_2(u). \label{eq:A4}
\end{eqnarray}
Here, $J_1$ and $J_2(q^2)$ denote the scalar Feynman integrals defined in
Eqs.~\eqref{tadpole_integral} and \eqref{eq:defJ2}, respectively. As in the previous cases, the on-shell condition $p_i^2 = M^2$ has been consistently applied.

Using dimensional regularization in $D=4-2\epsilon$ and renormalizing the low-energy constants according to
\begin{equation}
l_i=\frac{\gamma_i}{32\pi^2}
\left[
\bar{l}_i-\left(\frac1\epsilon+1+\ln\frac{\overline{\mu}^2}{M^2}\right)
\right],
\end{equation}
the final expression for the one-loop scattering amplitude becomes
\begin{eqnarray}
A(s,t,u) & = & \frac{s-M^2}{F^2} + \frac{1}{6F^4} \Bigg\{ \left[14M^4+t(s+2t)-2M^2(2s+5t) \right] \bar{J}_2(t) \nonumber \\
&& + \left[14M^4+u(s+2u)-2M^2(2s+5u) \right] \bar{J}_2(u) + 3(s^2-M^4) \bar{J}_2(s) \Bigg\} \nonumber \\
& & \mbox{}  + \frac{1}{96 \pi^2 F^4} \Bigg\{ 2 \bar{l}_1 \left(s-2 M^2\right)^2 +2 \bar{l}_2 \left(-8 M^4+4 M^2 s+t^2+u^2\right) \nonumber \\
& & \mbox{}  + \frac13 \left[53 M^4-24 M^2 s-8 s^2-5 \left(t^2+u^2\right)\right] \Bigg\}, \label{eq:Aoneloop}
\end{eqnarray}
where the function $\bar{J}(q^2)$ is defined as in Ref.~\cite{gasser},
\begin{equation}
\bar{J}(q^2) = \frac{1}{(4\pi)^2} \left[ 2 + \eta \ln\left(\frac{\eta-1}{\eta+1}\right) \right], \label{def_barJ2}
\end{equation}
with
\begin{equation}
\eta=\sqrt{1-\frac{4M^2}{q^2}}.
\end{equation}

The expression for $A(s,t,u)$ in Eq. \eqref{eq:Aoneloop} is in agreement with the classic result of Gasser and Leutwyler~\cite{gasser}.

\section{\label{app:FTidentityL24}Fourier transform identity for the 4-point functional derivative of the Lagrangian term $\mathcal{L}_2^{4\phi}$}

A common term that appears in the evaluation of the pion-pion scattering amplitude in Appendix~\ref{App:ScatteringAmplitudeOneloop} is the Fourier transform $\Gamma_{4 \phi}^{a_1 \dots a_4 }(p_1, \dots, p_4)$ of the 4-point functional derivative of $\mathcal{L}_2^{4\phi}$,
\begin{equation}
\Gamma_{4 \phi}^{a_1 \dots a_4 }(p_1, \dots, p_4) = \int d^D x_1 \dots d^D x_4\, \e^{i \sum_{i=1}^4 p_i \cdot x_i} \frac{\delta }{\delta \phi^{a_4}(x_4) \dots \delta \phi^{a_1}(x_1)} \int d^D x\, \mathcal{L}_2^{4\phi}.
\end{equation}

Using the expression for $\mathcal{L}_2^{4\phi}$ given in Eq.~\eqref{eq:Lcontorderphi}, the Fourier transform $\Gamma^{a_1 \dots a_4 }(p_1, \dots, p_4)$ can be rewritten as
\begin{equation}
\Gamma_{4 \phi}^{a_1 \dots a_4}(p_1,\dots,p_4) = (2\pi)^D \delta\left(\sum_{i=1}^4 p_i\right) \Lambda^{a_1a_2a_3a_4}(p1,\dots,p_4), \label{Gamma_4phi}
\end{equation}
where the Dirac-delta function ensures momentum conservation and the function $\Lambda^{a_1a_2a_3a_4}$ is defined as
\begin{eqnarray}
\Lambda^{a_1a_2a_3a_4}(p1,\dots,p_4) & = & \frac{1}{3F^2} \Bigg[ \Lambda^{(1)}(p_1, \dots, p_4) \delta^{a_1 a_4} \delta^{a_2 a_3} + \Lambda^{(2)}(p_1, \dots, p_4) \delta^{a_1 a_3} \delta^{a_2 a_4} \nonumber \\
& & \mbox{} + \Lambda^{(3)}(p_1, \dots, p_4) \delta^{a_1 a_2} \delta^{a_3 a_4} \Bigg], \label{def_Lambda}
\end{eqnarray}
with the functions $\Lambda^{(i)}$ given by,
\begin{subequations}
\label{Lambdas}
\begin{eqnarray}
\Lambda^{(1)} & = & M^2 - (p_1+p_4) \cdot (p_3+p_2) + 2 p_1 \cdot p_4 + 2 p_3 \cdot p_2, \\
\Lambda^{(2)} & = & M^2 - (p_1+p_3) \cdot (p_2+p_4) + 2 p_1 \cdot p_3 + 2 p_2 \cdot p_4, \\
\Lambda^{(3)} & = & M^2 - (p_1+p_2) \cdot (p_3+p_4) + 2 p_1 \cdot p_2 + 2 p_3 \cdot p_4.
\end{eqnarray}
\end{subequations}
Note that these functions are related to each other by permutations of the momenta $p_i$.

\section{ \label{app:feynmanint}One-loop integrals}

This appendix presents a set of identities for one-loop tensor and scalar Feynman integrals which arise in the evaluation of the pion-pion scattering amplitude and the pion self-energy.

The simplest case corresponds to the tadpole integral $J_1$ defined as
\begin{equation}
J_1 = i \int \frac{d^Dk}{(2\pi)^D} \frac{1}{k^2-M^2}, \label{tadpole_integral}
\end{equation}
which after using standard evaluation techniques, such as Feynman parametrization, becomes
\begin{equation}
J_1 = \frac{\Gamma\left(1-\frac{D}{2}\right)}{(4\pi)^{\frac{D}{2}}} (M^2)^{\frac{D}{2}-1}. \label{tadpole_int_eval}
\end{equation}

From the definition of $J_1$ in Eq.~\eqref{tadpole_integral}, and since scaleless integrals vanish in dimensional regularization \cite{abreu}, the following identity can be derived,
\begin{eqnarray}
i \int \frac{d^Dk}{(2\pi)^D} \frac{k^2}{k^2-M^2} & = & i \int \frac{d^Dk}{(2\pi)^D} \frac{(k^2-M^2) + M^2}{k^2-M^2} \nonumber \\
& = & M^2 J_1. \label{Iden_tadpole_int}
\end{eqnarray}

For a tensor integral with two scalar propagators, the tensor reduction prescription given in Ref.~\cite{denner} and the vanishing of scaleless integrals, yield the identities
\begin{subequations}
\begin{eqnarray}
i \int \frac{d^Dk}{(2\pi)^D} \frac{k^\mu}{(k^2-M^2)[(k+q)^2-M^2]} & = & - \frac{1}{2} q^\mu J_2(q^2), \\
i \int \frac{d^Dk}{(2\pi)^D} \frac{k^\mu k^\nu}{(k^2-M^2)[(k+q)^2-M^2]} & = & \frac{g^{\mu\nu}}{4(D-1)} \left[2J_1 + (4M^2-q^2) J_2(q^2)\right] \nonumber \\
& & \mbox{} + \frac{q^\mu q^\nu}{4q^2(D-1)} \left[2(D-2)J_1 + (Dq^2-4M^2)J_2(q^2)\right], \nonumber \\
\end{eqnarray}
\end{subequations}
where the two-point integral $J_2(q^2)$ is defined as,
\begin{equation}
J_2(q^2) = i \int \frac{d^Dk}{(2\pi)^D} \frac{1}{(k^2-M^2)[(k+q)^2-M^2]}. \label{eq:defJ2}
\end{equation}
After a standard calculation, the scalar integral $J_2(q^2)$ can be reexpressed as
\begin{equation}
J_2(q^2) = - \frac{\Gamma\left(2-\frac{D}{2}\right)}{(4\pi)^{\frac{D}{2}}} \int_0^1 \frac{dx}{[x(x-1)q^2+M^2]^{2-\frac{D}{2}}}. \label{eq:J2evalD}
\end{equation}

\section{\label{app:oneloopmr}One-loop corrections to the pion mass and to the residue of the exact pion propagator}

The physical mass of the pion $M_\pi$ is defined by the pole of the exact propagator of the pion at $p^2 = M_\pi^2$. In terms of the quantum effective action $\Gamma[\phi]$, this condition becomes,
\begin{equation}
\Gamma^{(2)}(M_\pi^2) = 0, \label{eq:masscond}
\end{equation}
where $\Gamma^{(2)}(p^2)$ is defined through the equation,
\begin{equation}
\int d^Dx_1 d^Dx_2 \, e^{i(p \cdot x_1 + q \cdot x_2)} \left.\frac{\delta \Gamma[\phi]}{\delta \phi^{a_2}(x_2) \delta \phi^{a_1}(x_1)}\right\vert_{\phi=0} = (2\pi)^D \delta(p+q) \delta^{a_1 a_2} \Gamma^{(2)}(p^2). \label{eq:defGamma2p}
\end{equation}

From Eq.~\eqref{eq:defQAoneloop_app}, for the quantum effective action at one-loop order, and Eq.~\eqref{eq:Logfunct_app}, the two-point functional derivative of $\Gamma[\phi]$ becomes,
\begin{eqnarray}
\left.\frac{\delta \Gamma[\phi]}{\delta \phi^{a_2}(x_2) \delta \phi^{a_1}(x_1)}\right\vert_{\phi=0} & = & \int d^Dx \, \frac{\delta (\mathcal{L}_2^{2\phi}+\mathcal{L}_4^{2\phi}) }{\delta \phi^{a_2}(x_2) \delta \phi^{a_1}(x_1)} - \frac{i}{2} \int d^D z_1 d^Dz_2 \, \Delta^{a^\prime_1 a^\prime_2}(z_1-z_2) \nonumber \\
& & \mbox{} \times \frac{\delta }{\delta \phi^{a_2}(x_2) \delta \phi^{a_1}(x_1)\delta \phi^{a^\prime_2}(z_2) \delta \phi^{a^\prime_1}(z_1)} \int d^D x\, \mathcal{L}_2^{4\phi}, \label{Two_point_fun_der}
\end{eqnarray}
where the Lagrangian terms $\mathcal{L}_2^{2\phi}$, $\mathcal{L}_2^{4\phi}$, and $\mathcal{L}_4^{2\phi}$ are defined in Eq.~\eqref{eq:Lcontorderphi}, and the function $\Delta^{ab}(x-y)$ is given by Eq.~\eqref{Prop_xspace}.

Now, from Eq.~\eqref{Gamma_4phi}, the functional derivative of the one-loop contribution on the right-hand-side of Eq.~\eqref{Two_point_fun_der} can be rewritten as,
\begin{eqnarray}
& & \int d^Dx\, \frac{\delta \mathcal{L}_2^{4\phi} }{\delta \phi^{a_2}(x_2) \delta \phi^{a_1}(x_1)\delta \phi^{a^\prime_2}(z_2) \delta \phi^{a^\prime_1}(z_1)} = (2\pi)^D \int \left(\prod_{i=1}^4\frac{d^Dk_i}{(2\pi)^D} \right) \frac{\delta(\sum_{i=1}^4 k_i)}{3F^2} \e^{-i w} \nonumber \\
& & \mbox{\hglue0.6truecm} \times \Bigg[ \Lambda^{(1)}(k_1, \dots, k_4) \delta^{a_1^\prime a_2} \delta^{a_2^\prime a_1} + \Lambda^{(2)}(k_1, \dots, k_4) \delta^{a_1^\prime a_1} \delta^{a_2^\prime a_2} + \Lambda^{(3)}(k_1, \dots, k_4) \delta^{a_1^\prime a_2^\prime} \delta^{a_1 a_2} \Bigg], \nonumber \\
\end{eqnarray}
where
\begin{equation}
w = k_1 \cdot z_1 + k_2 \cdot z_2 + k_3 \cdot x_1 + k_4 \cdot x_2.
\end{equation}

Thus,
\begin{eqnarray}
& & \int d^Dx_1 d^Dx_2 d^Dx \, e^{i(p \cdot x_1 + q \cdot x_2)} d^D z_1 d^Dz_2 \, \Delta^{a^\prime_1 a^\prime_2}(z_1-z_2) \frac{\delta \mathcal{L}_2^{4\phi} }{\delta \phi^{a_2}(x_2) \delta \phi^{a_1}(x_1)\delta \phi^{a^\prime_2}(z_2) \delta \phi^{a^\prime_1}(z_1)} \nonumber \\
& & \mbox{\hglue0.6truecm} = - (2\pi)^D \delta^{a_1 a_2} \frac{\delta(p+q)}{3F^2} \int \frac{dk}{(2\pi)^D} \frac{1}{k^2-M^2} \left[ \Lambda^{(1)} + \Lambda^{(2)} + 3 \Lambda^{(3)} \right]\vert_{\substack{k_1=k,k_2=-k\\ k_3=p,k_4=q}}, \label{one_loop_contribution_Gamma2}
\end{eqnarray}
where $\Lambda^{(i)}\vert_{\substack{k_1=k,k_2=-k\\ k_3=p,k_4=q}}$ gives the value of $\Lambda^{(i)}$ at $k_1=k$, $k_2=-k$, $k_3=p$ and $k_4=q$.

From the explicit expressions for $\Lambda^{(i)}$ of Eq.~\eqref{Lambdas}, the one-loop contribution in Eq.~\eqref{one_loop_contribution_Gamma2} reduces to
\begin{eqnarray}
& & \int d^Dx_1 d^Dx_2 d^Dx \, e^{i(p \cdot x_1 + q \cdot x_2)} d^D z_1 d^Dz_2 \, \Delta^{a^\prime_1 a^\prime_2}(z_1-z_2) \frac{\delta \mathcal{L}_2^{4\phi} }{\delta \phi^{a_2}(x_2) \delta \phi^{a_1}(x_1)\delta \phi^{a^\prime_2}(z_2) \delta \phi^{a^\prime_1}(z_1)} \nonumber \\
& & \mbox{\hglue0.6truecm} = (2\pi)^D \delta^{a_1 a_2} \frac{\delta(p+q)}{3F^2} \int \frac{dk}{(2\pi)^D} \frac{4p^2+4k^2-5M^2}{k^2-M^2}.
\end{eqnarray}

Thus, $\Gamma^{(2)}(p^2)$ becomes,
\begin{equation}
\Gamma^{(2)}(p^2) = p^2 - M^2 - \frac{2l_3 M^2}{F^2} - \frac{i}{6 F^2} \int \frac{d^Dk}{(2\pi)^D} \frac{4p^2+4k^2-5M^2}{k^2-M^2}.
\end{equation}

From Eqs.~\eqref{tadpole_integral}, \eqref{tadpole_int_eval}, and \eqref{Iden_tadpole_int}, setting $D=4-2\epsilon$, and introducing the renormalization mass scale parameter $\mu$ in such a way that $F$ remains to have the same mass-dimension for arbitrary value of $D$ as in the 4-dimensional case, $\Gamma^{(2)}(p^2)$ turns out to be
\begin{equation}
\Gamma^{(2)}(p^2) = p^2 - M^2 - \frac{2l_3 M^2}{F^2} + \frac{4p^2-M^2}{6 (4\pi)^2 F^2} M^2 \left[\frac{1}{\epsilon}+1+\ln\left(\frac{\overline{\mu}^2}{M^2}\right)\right], \label{Gamma2}
\end{equation}
where $\overline{\mu}^2 = 4 \pi e^{-\gamma_E} \mu^2 $ is the renormalization parameter in the $\overline{\text{MS}}$ scheme, with $\gamma_E$ being the Euler's constant.

Defining the variables $A$ and $B$ as
\begin{eqnarray}
A & = & - \frac{2l_3 M^2}{F^2} -\frac{M^4}{6 (4\pi)^2 F^2} \left[\frac{1}{\epsilon}+1+\ln\left(\frac{\overline{\mu}^2}{M^2}\right)\right], \\
B & = & \frac{M^2}{24 \pi^2 F^2} \left[\frac{1}{\epsilon}+1+\ln\left(\frac{\overline{\mu}^2}{M^2}\right)\right], \label{def_A_B_self_energy} 
\end{eqnarray}
the condition in Eq.~\eqref{eq:masscond} for the mass of the pion at order $\mathcal{O}(p^2)$ becomes,
\begin{eqnarray}
M_\pi^2 & = & M^2-B M^2-A \nonumber \\
& = & M^2 + \frac{2l_3 M^2}{F^2} - \frac{M^4}{32 \pi^2 F^2} \left[\frac{1}{\epsilon}+1+\ln\left(\frac{\overline{\mu}^2}{M^2}\right)\right]. \label{pion_mass_one_loop_condition}
\end{eqnarray}

The Lagrangian parameter $l_3$, borrowed from Ref.~\cite{gasser}, reads,
\begin{equation}
l_3 = l_3^{r}(\mu) + \frac{1}{64 \pi^2} \left[ \frac{1}{\epsilon} + \ln\left(\frac{\overline{\mu}^2}{\mu^2}\right)+1 \right], \label{def_l3}
\end{equation}
where $l_3^{r}(\mu)$ is the renormalized parameter that depends on the mass scale $\mu$,
\begin{equation}
l_3^{r}(\mu) = - \frac{1}{64 \pi^2} \left[ \overline{l}_3 + \ln\left( \frac{M^2}{\mu^2} \right) \right],
\end{equation}
so the expression for $M_\pi^2$ in Eq.~\eqref{pion_mass_one_loop_condition} reduces to the well-known one-loop chiral mass correction \cite{scherer1,gasser},
\begin{equation}
M_\pi^2 = M^2 \left( 1 - \frac{M^2}{32\pi^2 F^2} \overline{l}_3 \right). \label{eq:masspiononeloop}
\end{equation}
Here, $\overline{l}_3$ is a constant value independent of the mass scale $\mu$, which can be fixed by experiment.

Now, the residue $\mathcal{R}$ of the exact propagator of the pion at the simple pole $p^2 = M^2_{\pi}$ can be obtained from $\Gamma^{(2)}(p^2)$ through
\begin{equation}
\mathcal{R}^{-1} = \left.\frac{d}{dp^2} \Gamma^{(2)}(p^2)\right\vert_{p^2=M^2_\pi}. \label{def_residue}
\end{equation}

At one-loop order, the residue derived from Eqs.~\eqref{def_residue} and \eqref{Gamma2} reads,
\begin{equation}
\mathcal{R} = 1- \frac{2M^2}{3(4\pi)^2F^2}\left[\frac{1}{\epsilon}+1+\ln\left( \frac{\overline{\mu}^2}{M^2} \right)\right]. \label{Residue_oneloop}
\end{equation}

\section{\label{app:WavePack} Wave packet analysis }

Let $I$ be the integral that arises from the amplitude $M^{abcd}$ defined in Eq. \eqref{def:Mamplitude},
\begin{eqnarray}
    I = (2\pi)^4  \int \left( \prod_{i=1}^4 d^3 \bar{k}_i \right) \left( \prod_{i=1}^4 \phi_{\bm{p}_i}(\bm{k}_i) \right) \delta\left(\sum_{i=1}^4 k_i\right) A(k_i), \label{def:IntegralIsigma}
\end{eqnarray}
where $A(k_i)$ is a function of the momenta $k_i$. 

After performing the integral in $\bm{k}_4$, using the Dirac delta function for three-momentum, the expression for $I$ in Eq. \eqref{def:IntegralIsigma} becomes,
\begin{eqnarray}
    I = (2\pi)^4  \int \left( \prod_{i=1}^3 d^3 \bar{k}_i \right) \left( \prod_{i=1}^4 \phi_{\bm{p}_i}(\bm{k}_i) \right) \delta\left(\sum_{i=1}^4 E_{\bm{k}_i}\right) \dfrac{A(k_i)}{2E_{\bm{k}_4}}, \label{eq:IntegralIsigma_2} 
\end{eqnarray}
where $\bm{k}_4 = - \sum_{i=1}^3 \bm{k}_i$. 

Performing the change of variables $\bm{t}_i = (\bm{k}_i-\bm{p}_i)/\sigma$ for  $i=1,2,3$, the integral $I$ rewrite as,
\begin{eqnarray}
    I = (2\pi)^7 \sigma^3 \int \left( \prod_{i=1}^3 d^3 t_i \right) \delta\left(\sum_{i=1}^4 E_{\bm{k}_i}\right) \bar{A}(k_i) \e^{\mathcal{T}} \label{eq:IntegralIsigma_3}
\end{eqnarray}
with $\bm{k}_i = \bm{k}_i(\bm{t}_i)$ for $i=1,2,3$, $\bar{A}(k_i)$ defined as,
\begin{eqnarray}
    \bar{A}(k_i) = \left(\prod_{i=1}^4\dfrac{1}{\sqrt{2E_{\bm{k}_i}}}\right) A(k_i)
\end{eqnarray}
and the argument $\mathcal{T}$ of the exponential function given by
\begin{eqnarray}
    \mathcal{T} = -\dfrac{1}{4}\left[\sum_{i=1}^3 \bm{t}_i^2 + \left(\sum_{i=1}^3 \bm{t}\right)^2\right] - \dfrac{1}{2} \sum_{i=1}^3 \bm{t}_i \cdot \bm{P}_T - \dfrac{1}{4\sigma^2} \bm{P}_T^2. \label{def:mathcalT}
\end{eqnarray}
Here, $\bm{P}_T$ represents the total three-momentum of the peaks of the Gaussian wave packets,
\begin{eqnarray}
    \bm{P}_T = \sum_{i=1}^4 \bm{p}_i. \label{def:PTotal}
\end{eqnarray}
If $\bm{P}_T \neq \bm{0}$, the integral $I$ is suppressed by the exponential factor $\e^{- \frac{1}{4\sigma^2} \bm{P}_T^2}$, which vanishes in the limit when $\sigma \rightarrow 0$. Thus, to obtain a nonzero contribution, three-momentum conservation, $\bm{P}_T = 0$, must be satisfied at the peaks of the Gaussian profiles. Under this condition, the integral $I$ in the Eq. \eqref{eq:IntegralIsigma_3} becomes,
\begin{eqnarray}
    I = (2\pi)^7 \sigma^3 \int \left( \prod_{i=1}^3 d^3 t_i \right) dz \, \bar{A}(k_i) \; \e^{-\frac{1}{4}\left[\sum_{i=1}^3 \bm{t}_i^2 + \left(\sum_{i=1}^3 \bm{t}\right)^2\right]-iz \sum_{i=1}^4 E_{\bm{k}_i} }, \label{eq:IntegralIsigma_4}
\end{eqnarray}
where the energy conservation delta function has been rewritten in exponential form.  

The expression for the integral $I$ in Eq. \eqref{eq:IntegralIsigma_4} can be rewritten as a power series in $\sigma$ once the energy-conservation condition $\sum_{i=1}^4 E_{\bm{p}_i}=0$ is imposed \footnote{If this condition is not satisfied, the integral is exponentially suppressed by the factor $\e^{- \xi \left(\sum_{i=1}^4 E_{\bm{p}_i} \right)^2/\sigma^2 }$, where $\xi$ is a positive constant.}. This expansion is obtained by performing a Taylor series around the three-momenta $\bm{p}_i$, with $i=1,2,3$, of both $\bar{A}(\bm{k}_i)$ and $\sum_{i=1}^4 E_{\bm{p}_i}$ as functions of the variables $\bm{t}_i$.

Retaining only the leading order in $\sigma$, and after carrying out the Gaussian integrations with the rescaling $z=w/\sigma$, the integral $I$ in Eq. \eqref{eq:IntegralIsigma_4} reduces to,
\begin{eqnarray}
    I = \dfrac{(2\pi)^{11}}{\sqrt{\alpha\, \det \mathcal{Q} }}\, \bar{A}(\bm{p}_i)\; \sigma^2 + \mathcal{O}(\sigma^4), \label{eq:Isigma_result}
\end{eqnarray}
where 
\begin{eqnarray}
    \alpha &=& V^T \mathcal{Q}^{-1} V, \nonumber \\
    V^T &=& ( \bm{v}_1, \bm{v}_2, \bm{v}_3 ), \label{def:alphaV}
\end{eqnarray}
with $\bm{v}_i$ related to the velocities of the particles through the equation 
\begin{eqnarray}
    \bm{v}_i = \dfrac{\bm{p}_i}{E_{\bm{p}_i}} - \dfrac{\bm{p}_4}{E_{\bm{p}_4}},
\end{eqnarray}
and 
\begin{eqnarray}
    \mathcal{Q} = \begin{pmatrix}
        \mathds{1}_{3 \times 3} & \frac{1}{2} \mathds{1}_{3 \times 3} & \frac{1}{2} \mathds{1}_{3 \times 3} \\
        \frac{1}{2} \mathds{1}_{3 \times 3} & \mathds{1}_{3 \times 3} & \frac{1}{2} \mathds{1}_{3 \times 3} \\
        \frac{1}{2} \mathds{1}_{3 \times 3} & \frac{1}{2} \mathds{1}_{3 \times 3} & \mathds{1}_{3 \times 3}
    \end{pmatrix}. \label{def:mathcalQ}
\end{eqnarray}

\section{Von Neumman entropies for the basis states $\ket{I,M}$ } \label{app:SphericalBasis}

The flavor states of a two-pion composite system can be conveniently described in the basis $\ket{I,M}$, where $I$ denotes the total isospin of the two-pion system and $M$ its projection along the $z$-axis. In terms of cartesian basis, the states $\ket{I,M}$ can be expressed as,
\begin{eqnarray}
\ket{0,0} &=& \frac{1}{\sqrt{3}}
\left(
\ket{1,1}+\ket{2,2}+\ket{3,3}
\right), \nonumber \\
\ket{1,+1}
&=&
\frac{1}{2}
\Big[
(\ket{1,3}-\ket{3,1})
-i(\ket{2,3}-\ket{3,2})
\Big], \nonumber \\
\ket{1,0}
&=&
\frac{i}{\sqrt{2}}
\left(
\ket{1,2}-\ket{2,1}
\right), \nonumber \\
\ket{1,-1}
&=&
-\frac{1}{2}
\Big[
(\ket{1,3}-\ket{3,1})
+i(\ket{2,3}-\ket{3,2})
\Big],\nonumber \\
\ket{2,+2}
&=&
\frac{1}{2}
\Big[
(\ket{1,1}-\ket{2,2})
-i(\ket{1,2}+\ket{2,1})
\Big], \nonumber \\
\ket{2,+1}
&=&
\frac{1}{2}
\Big[
(\ket{1,3}+\ket{3,1})
-i(\ket{2,3}+\ket{3,2})
\Big], \nonumber \\
\ket{2,0}
&=&
\frac{1}{\sqrt{6}}
\Big(
\ket{1,1}+\ket{2,2}-2\ket{3,3}
\Big), \nonumber \\
\ket{2,-1}
&=&
-\frac{1}{2}
\Big[
(\ket{1,3}+\ket{3,1})
+i(\ket{2,3}+\ket{3,2})
\Big], \nonumber \\
\ket{2,-2}
&=&
\frac{1}{2}
\Big[
(\ket{1,1}-\ket{2,2})
+i(\ket{1,2}+\ket{2,1})
\Big]. \label{def:sphericalbasis}
\end{eqnarray}

The degree of entanglement of the states $\ket{I,M}$ can be quantified by computing the von Neumann entropy of the corresponding reduced density matrix $\rho^{(I,M)}$, obtained by tracing over one of the pion flavor degrees of freedom. This matrix is defined as
\begin{eqnarray}
    \rho^{(I,M)} = \sum_a \braket{a|I,M}\braket{I,M|a},
\end{eqnarray}
which, when expressed in the cartesian basis, take the form of $3\times3$ matrices given by,
\begin{eqnarray}
    \rho^{(0,0)} &=& \left(
\begin{array}{ccc}
 \frac{1}{3} & 0 & 0 \\
 0 & \frac{1}{3} & 0 \\
 0 & 0 & \frac{1}{3} \\
\end{array}
\right), \quad 
\rho^{(1,1)} = \left(
\begin{array}{ccc}
 \frac{1}{4} & \frac{i}{4} & 0 \\
 -\frac{i}{4} & \frac{1}{4} & 0 \\
 0 & 0 & \frac{1}{2} \\
\end{array}
\right), \quad 
\rho^{(1,0)} = \left(
\begin{array}{ccc}
 \frac{1}{2} & 0 & 0 \\
 0 & \frac{1}{2} & 0 \\
 0 & 0 & 0 \\
\end{array}
\right), \nonumber \\
\rho^{(1,-1)} &=& \left(
\begin{array}{ccc}
 \frac{1}{4} & -\frac{i}{4} & 0 \\
 \frac{i}{4} & \frac{1}{4} & 0 \\
 0 & 0 & \frac{1}{2} \\
\end{array}
\right), \quad 
\rho^{(2,2)} = \left(
\begin{array}{ccc}
 \frac{1}{2} & \frac{i}{2} & 0 \\
 -\frac{i}{2} & \frac{1}{2} & 0 \\
 0 & 0 & 0 \\
\end{array}
\right), \quad 
\rho^{(2,1)} = \left(
\begin{array}{ccc}
 \frac{1}{4} & \frac{i}{4} & 0 \\
 -\frac{i}{4} & \frac{1}{4} & 0 \\
 0 & 0 & \frac{1}{2} \\
\end{array}
\right), \nonumber \\
\rho^{(2,0)} &=& \left(
\begin{array}{ccc}
 \frac{1}{6} & 0 & 0 \\
 0 & \frac{1}{6} & 0 \\
 0 & 0 & \frac{2}{3} \\
\end{array}
\right), \quad 
\rho^{(2,-1)} = \left(
\begin{array}{ccc}
 \frac{1}{4} & \frac{i}{4} & 0 \\
 -\frac{i}{4} & \frac{1}{4} & 0 \\
 0 & 0 & \frac{1}{2} \\
\end{array}
\right), \quad 
\rho^{(2,-2)} = \left(
\begin{array}{ccc}
 \frac{1}{2} & -\frac{i}{2} & 0 \\
 \frac{i}{2} & \frac{1}{2} & 0 \\
 0 & 0 & 0 \\
\end{array}
\right).  \nonumber \\ \label{eq:redmatspherical}
\end{eqnarray}

Using Eqs.~\eqref{eq:VNentropyeigen} and \eqref{eq:redmatspherical}, the von Neumann entropies associated with the basis states $\ket{I,M}$ are summarized in Table~\ref{table:SphericalBasisVNE}.

\begin{table}[h!]
  \centering 
  \begin{tabular}{|l|c|}
    \hline
    States & Von Numann entropy  \\ \hline
    $\ket{0,0}$ (Maximally entangled qutrit)   & $\log 3$      \\ \hline
    $\ket{1,M}$ (Maximally entangled qubit)   & $1$      \\ \hline   
    $\ket{2,2}$ (Unentangled state)   & 0      \\ \hline     
    $\ket{2,1}$ (Maximally entangled qubit)   & 1      \\ \hline         
    $\ket{2,0}$ (Entangled qutrit)   & $\log 3-\frac{1}{3}$      \\ \hline             
  \end{tabular}
  \caption{Von Neumann entropies for the basis states $\ket{I,M}$} 
  \label{table:SphericalBasisVNE}
\end{table}

\end{document}